\documentclass{emulateapj}
\usepackage{graphics}

\def\msun{\,  {\rm M_\odot}}
\def\cm-3{\,{\rm cm^{-3}}}

\def\kpc-3{\,{\rm kpc^{-3}}}
\def\myr-1{\,{\rm Myr^{-1}}}
\def\pc{\,{\rm pc}}
\def\kpc{\,{\rm kpc}}

\def\hSN{$h_{\rm{SN,cc}}$}
\def\hgas{$h_{\rm{gas}}$}
\def\ml{$\eta_{\rm{m}}\ $}
\def\el{$\eta_{\rm{E}}\ $}
\def\mel{$\eta_{\rm{met}}\ $}
\def\siggas{$\Sigma_{\rm{gas}}\ $}

\def\fSN{$f_{\rm{SN, h}}\ $}
\def\vB{$v_\mathcal{\widetilde{B}}\ $}
\def\sigSFR{$\dot{\Sigma}_{\rm{SFR}}$}

\usepackage{natbib}
\usepackage{color}
\usepackage{rotating}
\usepackage{footnote}
\usepackage{datetime}
\usepackage{amsmath}
\usepackage{mathrsfs}
\usepackage[normalem]{ulem}

\begin{document}

\title{Quantifying Supernovae-Driven Multiphase Galactic Outflows}

\author{Miao Li\altaffilmark{1,$\dagger$}, Greg L. Bryan\altaffilmark{1,2}, Jeremiah P. Ostriker\altaffilmark{1}} 

\altaffiltext{$\dagger$}{miao@astro.columbia.edu}
\altaffiltext{1}{Department of Astronomy, Columbia University, 550 W120th Street, New York, NY 10027, USA} 
\altaffiltext{2}{Simons Center for Computational Astrophysics, 160 5th Ave, New York, NY, 10010, USA}

\begin{abstract}
{
Galactic outflows are ubiquitously observed in star-forming disk galaxies and are critical for galaxy formation. Supernovae (SNe) play the key role in driving the outflows, but there is no consensus as to how much energy, mass and metal they can launch out of the disk. We perform 3D, high-resolution hydrodynamic simulations to study SNe-driven outflows from stratified media. Assuming SN rate scales with gas surface density $\Sigma_{\rm{gas}}$ as in the Kennicutt-Schmidt (KS) relation, we find the mass loading factor, \ml, defined as the mass outflow flux divided by the star formation surface density, decreases with increasing \siggas as \ml$\propto \Sigma^{-0.61}_{\rm{gas}}$. Approximately $\Sigma_{\rm{gas}} \lesssim 50 M_\odot /\pc^2$ marks when \ml $\gtrsim$1. About 10-50\% of the energy and 40-80\% of the metals produced by SNe end up in the outflows. The tenuous hot phase ($T>3\times 10^5$ K), which fills 60-80\% of the volume at mid-plane, carries the majority of the energy and metals in outflows. We discuss how various physical processes, including vertical distribution of SNe, photoelectric heating, external gravitational field and SN rate, affect the loading efficiencies. The relative scale height of gas and SNe is a very important factor in determining the loading efficiencies.

 }

\end{abstract}

\section{Introduction}
\label{intro}

Galactic outflows, widely observed in star forming galaxies, share a few universal properties. First, outflows are multi-phase. The nearby starburst galaxy M82 is a well-studied example: the outflowing gas consists of a hot phase emitting X-rays, a warm ionized phase probed by H$\alpha$, and a cool, dusty phase seen in the infrared. At higher redshifts $z \sim 1-3$, warm/cool outflows have been widely observed in emission and absorption \citep[e.g.][]{steidel96, shapley03, martin05, weiner09, chen10, genzel11, heckman15}. Some even contain molecular components \citep[e.g.][]{walter02,bolatto13}. Hot winds have also been detected recently \citep{turner15}. 

Second, the velocities of outflows are several hundred km/s. This is comparable to the escape velocities from dark matter halos, indicating that outflows strongly impact galactic evolution, the circumgalactic medium (CGM) and even the intergalactic medium (IGM). 

Third, the mass loading factor, defined as the ratio between the outgoing mass flux and the star formation rate (SFR), ranges from 0.01-10 \citep[see review by][and references therein]{veilleux05}. For star-burst systems, warm/cool outflows have commonly been reported to have mass loading $\gtrsim$1. But note that large uncertainties in this quantity remain, since the geometry, metallicity, and ionization fraction of the outflows are hard to constrain, and a smaller loading factor  ($\sim 10\%$) is possible \citep{chisholm16}.

Galactic outflows play a critical role in galaxy formation. Without them, galaxies in cosmological simulations becomes too massive, too small, and too metal-enriched \citep[e.g.][]{scannapieco08}. Outflows remove gas from galaxies and delay gas infall, thus limiting a galaxy's mass \citep[e.g.][]{springel03}. Some ejected mass may eventually fall back at the edge of the galaxy, building the disk from the inside out \citep{governato07, genel15}. Outflows also funnel metals from galaxies to their surroundings\citep[e.g.][]{maclow99, fujita04, oppenheimer06}. This naturally explains why galaxies retain less metals than they have produced \citep{tremonti04,erb06}, while the CGM and IGM are metal-enriched \citep{mitchell76, songaila96, schaye03}.

Supernovae (SNe) explosions, the most energetic process associated with stellar feedback, are thought to be the main driving force of the outflows for galaxies with $M\lesssim 10^{10} \msun$ \citep{efstathiou00}. The general picture is that supernova remnants (SNRs) overlap and create hot bubbles, which break out of the disk and launch outflows \citep{cox74, mckee77,cox05}. But under what conditions can SNRs overlap? How much energy, mass and metals are carried in the outflows? Can the outflows escape the galaxy? The answers are essential not only for understanding the observations, but also for building a physically-based model of feedback for cosmological simulations and semi-analytic models of galaxy formation. Ad hoc feedback models have been widely used in those works, thus the predictive power is severely limited \citep[see recent reviews by][Naab \& Ostriker 2016]{somerville15}. 

High resolution numerical simulations are essential to study SN feedback, given the complexities of the multiphase ISM and the non-linear interactions of blast waves. \cite{li15} present a high-resolution study to show how SNe shape a patch of the ISM under various conditions,. They find, for a given mean gas density, the critical SN rate for hot bubbles to overlap, and to produce a multiphase medium. Various papers have explored how SNe drive outflows from a stratified medium \citep[e.g.][]{deavillez00, joung06, joung09, hill12, gent13, walch15}. The solar neighbourhood is the most well-studied case, in which the mass loading factor of order unity is found. \cite{creasey13} explored a wide parameter space of gas surface density and external gravitational field. Assuming the SN rate correlates with the gas density via the empirical star formation law --  the Kenniccut-Schmidt (KS) relation -- they found that the mass loading decreases with increasing gas surface density.

In this paper, we study SNe-driven outflows from a stratified disk, with various gas surface densities and SN rates. We focus on the following questions: 
(1) How much energy, mass and metals can SNe launch out of the disk? How do these change along the Kennicutt-Schmidt relation? 
(2) What physical processes affect the energy, mass and metal loading?  We explore the effects of runaway OB stars, photoelectric heating (PEH), gravitational field, enhanced SN rates, etc.
While our simulations focus on regions around the disk ($\pm$ 2.5 kpc), we discuss how our results connect to outflows on galactic scale.

We organize our paper as follows: we describe the numerical set-ups in Section 2, present the results of the fiducial models in Section 3, and discuss the various processes that can affect the loading efficiencies in Section 4. We discuss the implications of our results in Section 5, and summarize in Section 6.

\section{Methods}
\label{sec:method}

\subsection{Simulation Set-ups}

The simulations are performed using the Eulerian hydrodynamical code \textsc{Enzo}  \citep{bryan14}. We set up a rectangular box with z-dimension of 5 kpc ($-2.5 \leqslant z \leqslant 2.5$ kpc). The midplane of the disk is located at $z=0$. The horizontal cross section, i.e., x-y plane, is a square. The length of the horizontal dimension, $l_x$, varies with $\Sigma_{\rm{gas}}$, as listed in Table 1. The idea is that we adopt higher resolutions for larger \siggas, while keeping the corresponding $l_x$ smaller to gain computational efficiency (but sufficiently large to include many SNRs). The grid is refined near the midplane, with two refinement levels. Each refinement increases the resolution by a factor of two. The first level is within $|z|\leqslant$ 1 kpc, and the second is $|z|\leqslant$ 0.5 kpc. The finest resolution for each run is so chosen that the cooling radius of a SNR $R_{\rm{cool}}$ is resolved by approximately 12 cells for the initial midplane density $\rho_{\rm{mid}}$. (For the definition of $R_{\rm{cool}}$, see Eq. 1 of Li et al. 2015.) \cite{kim15} have shown that resolving $R_{\rm{cool}}$ by 10 cells is necessary to well-capture the evolution of a SNR in the Sedov-Taylor phase. Once the ISM becomes multiphase, SNe exploding in the dense region could be under-resolved. But as we discuss in Section \ref{sec:miss_phy}, this is likely a minor issue.

The boundary conditions are periodic for the x- and y-directions, and outflowing for z. We use the finite-volume piece-wise parabolic method \citep{colella84} as the hydro-solver. We use the cooling curve as \cite{rosen95},  for the temperature range of $300-10^9$ K.  Photoelectric heating (PEH) is time-independent and uniform across the box. The rate of PEH scales linearly with the star formation surface density \sigSFR; for the solar neighbourhood model $\Sigma$10-KS (see below), we adopt a PEH rate of $1.4 \times 10^{-26} \rm {erg\ s^{-1} }$ per H atom \citep{draine11}. We explore the variations of the PEH that deviate from the fiducial settings in Section \ref{sec:PEH}. 

\begin{figure}
\begin{center}
\includegraphics[width=0.47\textwidth]{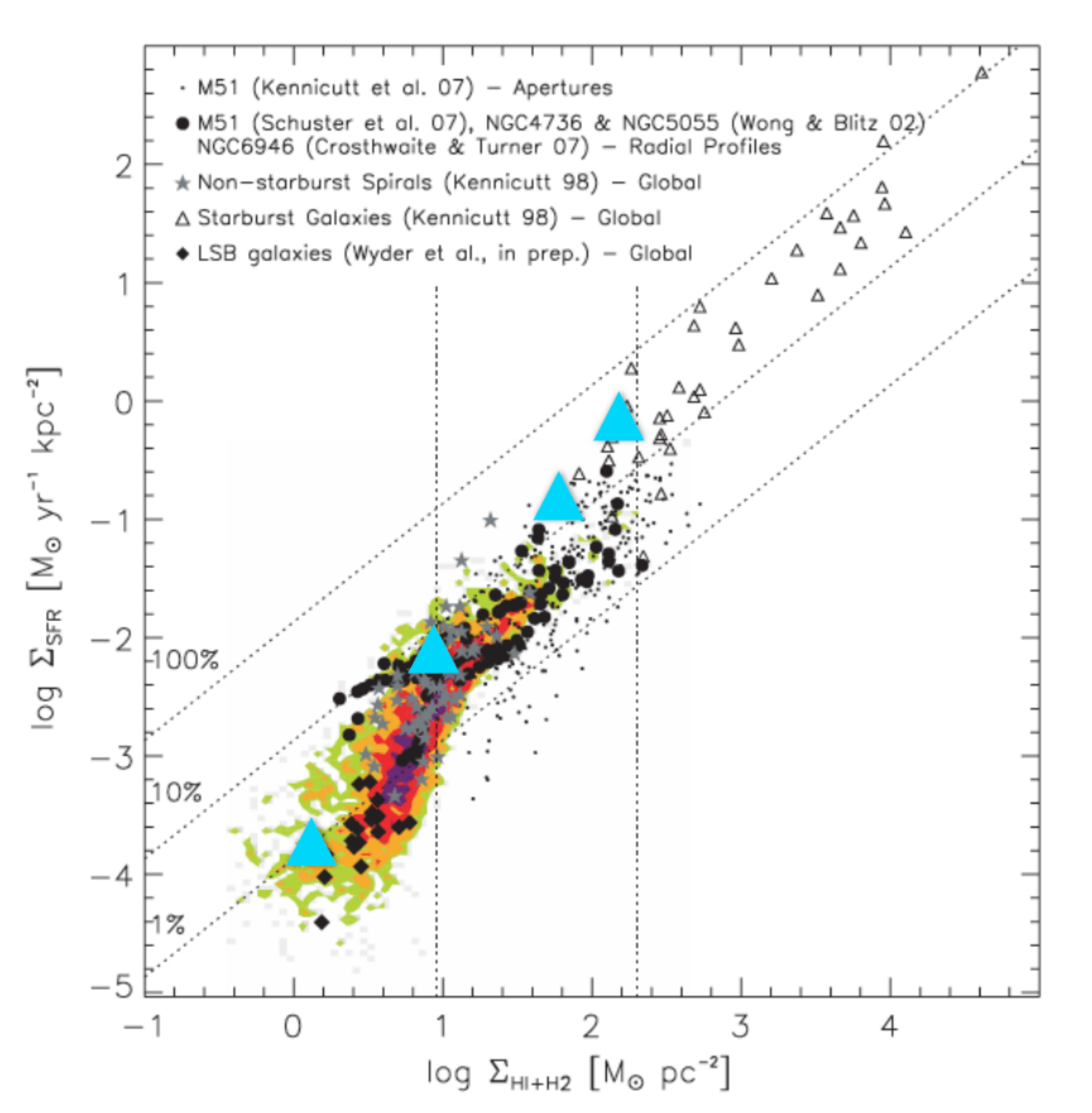}
\caption{Combinations of \siggas and $\dot{\Sigma}_{\rm{SFR}}$ adopted in our simulations (blue triangles), plotted on top of Fig 15 of \cite{bigiel08}, which shows the observed correlations for nearby galaxies at sub-kpc scale. }
\label{f:bigiel08}
\end{center}
\end{figure} 

We have four fiducial runs: $\Sigma1$-KS, $\Sigma10$-KS, $\Sigma55$-KS, $\Sigma150$-KS. The number after $\Sigma$ indicates the gas surface density in units of $\msun/\rm{pc}^2$. The SFRs associated with those runs are along the KS relation. Fig \ref{f:bigiel08} shows the (\siggas, $\dot{\Sigma}_{\rm{SFR}}$) adopted in our simulations, indicated by blue triangles. They are plotted on top of Fig 15 of \cite{bigiel08}, which shows the observed correlations of \siggas and $\dot{\Sigma}_{\rm{SFR}}$ for nearby galaxies on sub-kpc scales. Table \ref{t:runs} summarizes the setups of the simulations. $\Sigma$10-KS is the model for the solar neighbourhood. Variations of fiducial runs are described in Section \ref{sec:multi-effect}. The gravitational field, initial gas distribution and the model of SN feedback are detailed in the next two sub-sections. For the fiducial run $\Sigma10$-KS, we have carried out a resolution convergence check, where we lower the spatial resolution by a factor of 2. The results agree very well, including the ISM properties, volume fraction of difference gas phases, and outflow fluxes.

\subsection{Gravitational fields}
\label{sec:intro_g}

The gravitational field (``g-field" hereafter) has two components: a baryonic disk and a dark matter (DM) halo. The disk is modeled as self-gravitating with an iso-thermal velocity dispersion, so its g-field has the form $g = 2\pi G \Sigma_{*} \ \rm{tanh} (z/z_*)$, where $z_*$ is the scale height of the stellar disk: $z_*\equiv \sigma_*^2/(\pi G \Sigma_*)$, in which $\sigma_*$ and $\Sigma_*$ are the velocity dispersion and the surface density of stars, respectively. The height $z_* = $ 300 pc is observed for the solar neighbourhood \citep{gilmore83,binney08}; we keep it unchanged for all our runs. Since we do not include self-gravity of the gas, we multiply the stellar gravitational field by a factor of $1/f_*$, where $f_* = \Sigma_*/(\Sigma_* +\Sigma_{\rm{gas}})$. The g-field from the baryonic disk is
\begin{equation}
g_{\rm{disk}} (z) = \frac{1}{f_*} 2\pi G \Sigma_{*} \ \rm{tanh} (\frac{z}{z_*}),
\end{equation}

The g-field from the DM halo is modeled as an NFW profile projected to the z-direction,
\begin{equation}
g_{\rm{DM}}(z) = \frac{GM_{\rm{DM}}(r)\ z}{r^3},
\end{equation}
where $M_{\rm{DM}}(r)$ is the enclosed mass of a NFW halo within radius $r$, so
$M_{\rm{DM}}(r) = 4\pi \rho_{\rm{DM}} R_s^3\  \{ \rm{ln}(1+r/R_s) - r/(r+R_s) \}$, $R_s = R_{\rm{vir}}/c$, $\rho_{\rm{DM}} = 200\ \bar{\rho}_{\rm{DM}}\ c(1+c)^2$; $\bar{\rho}_{\rm{DM}}$ is the mean cosmic DM density at redshift 0, and c is the concentration parameter. For the MW case, we take $R_{\rm{vir}} = $ 200 kpc, and $c=12$ \citep{navarro97}. Note that $r$ and $z$ are related through $r^2=z^2+R_d^2$, in which $R_d$ is the distance from the location of the ISM patch we simulate, to the galactic center.
The total gravitational field is therefore
\begin{equation}
g_{\rm{tot}} =  g_{\rm{disk}} + g_{\rm{DM}}.
\end{equation}
Table 1 lists $\Sigma_*$ for all of our runs. We keep $R_d=8$ kpc for all simulations, except for $\Sigma10$-KS-4g which has $R_d=3$ kpc (see Section \ref{sec:g-field} for details). In Table 1 we also include $v_{\rm{\Delta \phi}}$ (see table footnote for its definition) as an indicator for the total potential well for each simulation box. 

\begin{figure}
\begin{center}
\includegraphics[width=0.5\textwidth]{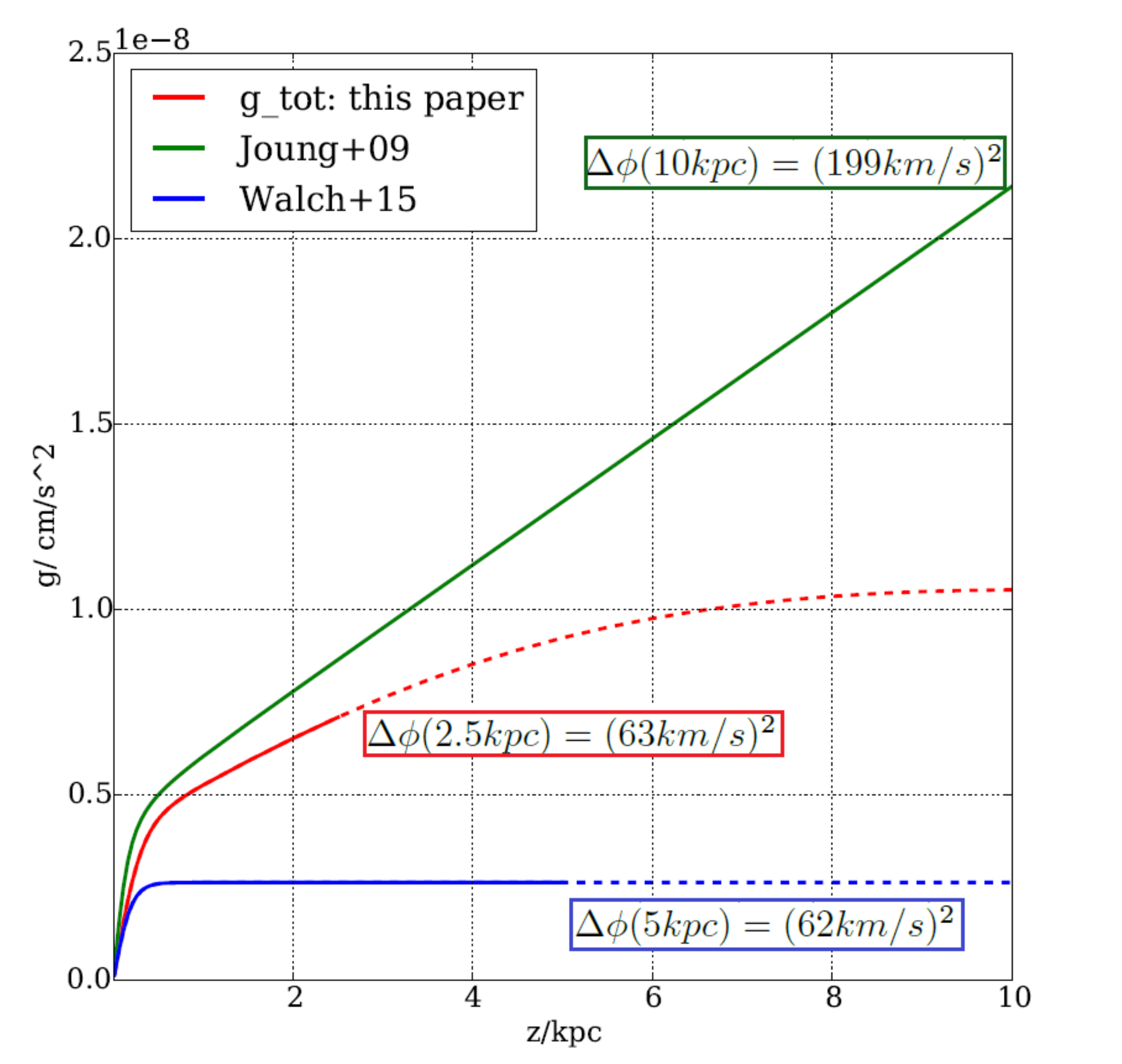}
\caption{Comparison of g-fields adopted in literature and this work for the solar neighbourhood. The solid lines end at $z_{\rm{max}}$ of each simulation box, which are 2.5kpc, 5kpc and 10kpc for this work, \cite{walch15} and \cite{joung09}, respectively. $\Delta\phi$ for each curve shows the gravitational potential $\Delta\phi (z_{max}) = \int_{z=0}^{z_{max}} g_{\rm{tot}} dz $. See Section \ref{sec:intro_g} for details. }
\label{f:g_compare}
\end{center}
\end{figure}

Note that in literature the adopted g-field can vary by a factor of a few even when the same ``solar neighbourhood'' is claimed. In Fig \ref{f:g_compare} we compare our value to a few others. \cite{walch15} do not include the DM halo potential, so they have a smaller $g$; at $z=5$ kpc, their g-field is about $1/3$ of our value. \cite{joung09} uses the observed g-field in the solar neighbourhood from \cite{kuijken89}, and extrapolates it into the halo. This works for $z\lesssim$ 1-2 kpc, but above that a simple extrapolation is likely too large. $\Delta\phi$ for each curve shows the gravitational potential $\Delta\phi (z_{\rm{max}}) = \int_{z=0}^{z_{\rm{max}}} g_{\rm{tot}} dz $. The numerical values of $\Delta\phi$ are not negligible compared to the kinetic energies of the outflows, which are typically 100-500 km/s (see Section \ref{sec:vel} below). Consequently, the gravitational field is dynamically important. Indeed, the value of g-field turns out to be important for the loading efficiencies of the outflows (Section \ref{sec:load}). Any meaningful comparisons between different works should take into account the difference in g-fields.

Initially the gas has a uniform temperature  $T_{0} = 10^4$ K. 
We set up the gas initial density to be in hydro-static equilibrium in the g-field $g_{\rm{disk}}(z)$, i.e.,
\begin{equation}
\rho = \rho_{\rm{mid}} \ \{\rm{sech}(\frac{z}{z_*})\}^{2\alpha},
\end{equation}
where $\alpha= \gamma \sigma_{*}^2 /(f_* c_{s,0}^2)$, $\gamma = 5/3$ is the adiabatic index of the gas and $c_{s,0}$ is the sound speed for $T_0$. 
The power-law decay at large z can result in very low density, so we set up a density floor of $3\times 10^{-28}\ \rm{g\ cm^{-3}}$.
Due to the density floor and an extra gravitational field from the DM, the gas is not in perfect hydrostatic equilibrium, but in practice this has little consequence, since the outflowing gas will soon dominate the space above the mid-plane.

\subsection{SN Feedback models}
\label{sec:SN_model}

The SN surface density is related to $\dot{\Sigma}_{\rm{SFR}}$ by assuming one SN explosion per $m_0=$150 $\msun$ star formation. There are some uncertainties associated with $m_0$; different works have adopted $m_0=$ 100-200 $\msun$. The distribution of SNe over time is uniform.
SNe are randomly located horizontally; in the z-direction, the distribution is stratified. We distinguish two components of SNe: Type Ia and core collapse SNe. Type Ia constitutes 10\% of SNe occurrence and core collapse the rest. 
Type Ia SNe have an exponential distribution in z-direction, with a scale height of 325 pc, similar to the old stellar disk \citep{freeman87}. Core collapse SNe have a Gaussian distribution vertically with a scale height $h_{\rm{SN,cc}} =$ 150 pc. We note that due to runaway OB stars, core collapse SNe may explode outside of the dense gas layer. We test the sensitivity of the results on $h_{\rm{SN,cc}}$, as described in Section \ref{sec:SN_height}.

Each SN is implemented as injecting $E_{\rm{SN}}=10^{51}$ erg energy, $m_{\rm{SN}}=$ 10 $\msun$ mass, and $m_{\rm{0,met}}$ metals (metals are modeled as ``color tracers'' that passively follow the mass flux, in arbitrary units), evenly distributed in a sphere. The energy added is 100\% thermal. The injection radius $R_{\rm{inj}} $ varies for \siggas, and chosen to be 0.45-0.50 of the cooling radius for the initial midplane density $\rho_{\rm{mid}}$. \cite{kim15} argued that $R_{\rm{inj}}/R_{\rm{cool}}\leqslant 1/3$ is the robust criterion to capture the evolution of a SNR in the Sedov-Taylor phase. Our choice is slightly larger than that. 

%\begin{sidewaystable} 
\begin{center}
\begin{table*}

\caption{Model description}

  \begin{tabular}{c | c    c    c  c  c  c   c   c  | c c c  c } 
    \hline
Run
&  \begin{tabular}{@{}c@{}} \siggas \\ ($\msun/\rm{pc}^2$) \end{tabular} 
& \begin{tabular}{@{}c@{}} $\rho_{\rm{mid}}$\\ (cm$^{-3}$) \end{tabular}
& \begin{tabular}{@{}c@{}} $\dot{\Sigma}_{\rm{SFR}}$ \footnote{SN rate is $\dot{\Sigma}_{\rm{SFR}}$/$m_0$, where $m_0 = 150\msun$} \\ ($\msun/\rm{kpc^2/yr}$) \end{tabular} 
&\begin{tabular}{@{}c@{}} PEH\\ (erg/s)\end{tabular} 
& \begin{tabular}{@{}c@{}} $\Sigma_*$ \\ (M$_\odot$) \end{tabular} 
& \begin{tabular}{@{}c@{}} $h_{\rm{SN,cc}}$  \\(pc)\end{tabular}  
& \begin{tabular}{@{}c@{}} $v_{\rm{\Delta \phi}}$ \footnote{$v_{\rm{\Delta \phi}} ^2 \equiv 2 \Delta \phi (z_{max})= 2\int_{z=0}^{z_{max}} g_{\rm{tot}} dz $ } \\ (km/s) \end{tabular} 
& \begin{tabular}{@{}c@{}} $T_{\rm{min}}$\footnote{temperature cut-off of the cooling curve} \\ (K) \end{tabular}   
& \begin{tabular}{@{}c@{}} Res \footnote{resolution at $z\leqslant$500 pc}\\(pc)\end{tabular}
 & \begin{tabular}{@{}c@{}}$l_x$ \\ (pc) \end{tabular}
 & \begin{tabular}{@{}c@{}}$R_{\rm{inj}}$ \\ (pc) \end{tabular}
 & \begin{tabular}{@{}c@{}} $t_{\rm{sim}}$\footnote{simulation time} \\ (Myr) \end{tabular}
\\ \hline

     $\Sigma 1$-KS &  1 &0.011 & 1.26E-4 & 2.8E-28 & 0.5  & 150  & 52 &300  & 12.5 &1200 & 80.0 & 300 \\
     $\Sigma 1$-3KS  & .. &.. & 3.78E-4 & .. & .. & .. & ..&.. &.. &.. &.. & 240  \\
     $\Sigma 1$-10KS & .. &.. & 1.26E-3 & .. & .. & .. & ..&.. &.. & .. &..  & 130 \\\hline
     
     $\Sigma 10$-KS &  10 & 0.822 & 6.31E-3 & 1.4E-26 & 35 & .. & 89 & ..&  2.0 & 350 & 12.0  & 160 \\   
     $\Sigma 10$-KS-4g &  .. & 1.78 & .. & .. & 180 & .. & 178 &.. & .. &..  & .. & 64\\   \hline

     $\Sigma 55$-KS-h75 &  55 & 8.2 & 0.158 & 3.5E-25 & 35  &75 & 117 &.. & 0.75& 150 & 4.5 & 40  \\
     $\Sigma 55$-KS(-h150) &  .. & .. & .. & .. & .. & 150 &..& .. & .. & .. & ..&.. \\
     $\Sigma 55$-KS-h300&  .. & .. & .. & .. & ..  & 300&..& ..  & .. & .. & ..&.. \\
     $\Sigma 55$-KS-h450 &  .. & .. & .. & .. & ..  & 450 &..& .. & .. & ..& .. &.. \\ \hline
     $\Sigma 55$-KS-noPEH &  .. & .. & .. &0 & ..  & 150 &.. & .. & .. & .. & .. &.. \\
     $\Sigma 55$-KS-5PEH &  .. & .. & .. & 1.75E-24& .. & .. & .. &..& .. & ..& .. &.. \\ 
      $\Sigma 55$-KS-1e4K &  .. & .. & .. &  0 & .. & .. & .. & $10^4$  & .. & .. & .. &.. \\  \hline

     $\Sigma 150$-KS &  150 & 26.3 & 0.708 & 1.6E-24 & .. & ..  & 160  & 300  & 0.60 & 122 & 3.0 & 15 \\\hline

  \end{tabular}
  \label{t:runs}
  \end{table*}
    \end{center}

%  \end{sidewaystable}

\section{Results of fiducial runs}

Impacted by SN explosions, the stratified medium quickly becomes multiphase, in which the hot gas occupies a significant fraction of the volume while most mass is in cooler clouds. Cool gas settle down to near the midplane, while outflows are launched. In this section, we first examine the multiphase structure of the ISM, with the emphasis of the comparison between our solar neighbourhood run with the observations. Then we discuss the velocities of the outflows, and show that the hot phase has the strongest potential to travel to large radii in the DM halo and impact the CGM. Finally we show the mass, energy and metal loading factors as a function of \siggas. 

\subsection{Multiphase ISM and outflows}
\label{sec:multiphase}

\begin{figure}
\begin{center}
\includegraphics[width=0.5\textwidth]{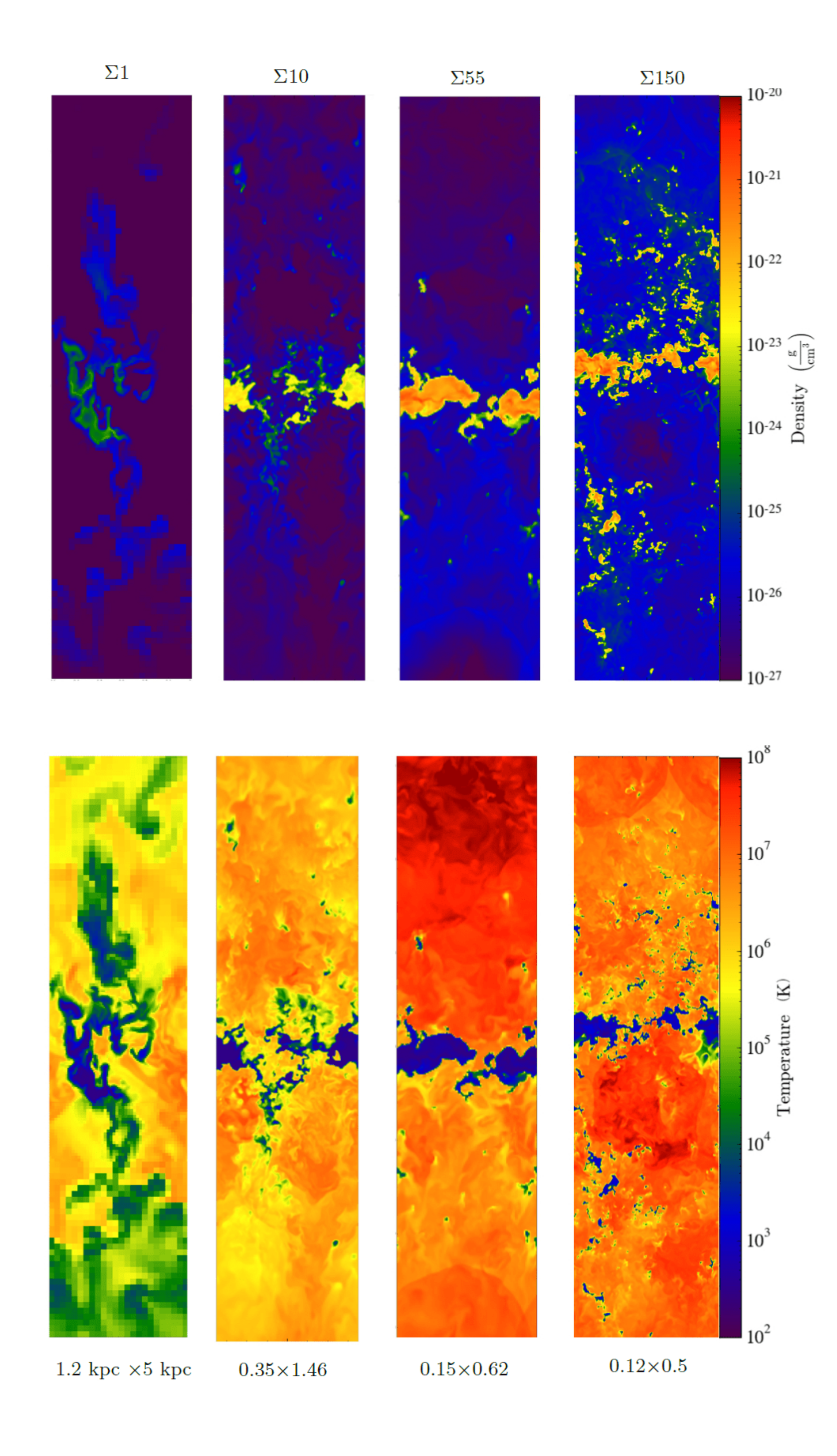}
\caption{Density and temperature slices of the x-z plane for the four fiducial runs.
Note that the physical scales of the slices are different. The dimensions are shown at the bottom of the temperature slices, in format of ``horizontal scale $\times$ vertical", in units of kpc. }
\label{f:slice_all}
\end{center}
\end{figure} 

Fig \ref{f:slice_all} shows the slices of the four fiducial runs in the x-z plane. Note that the physical scales of the slices are different from each other. The horizontal lengths are same as the simulation boxes; the vertical dimensions are shown partially. The actual dimensions each slice represents are indicated at the bottom of the temperature slices. 

Most of the dense gas stays near the midplane. The medium has multiple phases -- a cold phase at a few hundred K, a warm phase at around $10^4$ K and a hot phase at $T\gtrsim 10^6$ K. At the boundary between the hot and warm/cold phase, gas with intermediate temperature,  $10^{5-6}$ K, is also seen. For all four runs, the hot gas volume fraction is about 60-80\% for the midplane. Hot gas occupies more volume in the halo for higher \siggas .  Multiphase outflows are being launched from the midplane for all four runs.  Cool clouds in the halo are clearly being stripped by the hot, faster gas. The hot phase appears hotter for higher \siggas . These qualitative results agree with previous works \citep{mckee77, joung09, creasey13}.

\begin{figure}
\begin{center}
\includegraphics[width=0.5\textwidth]{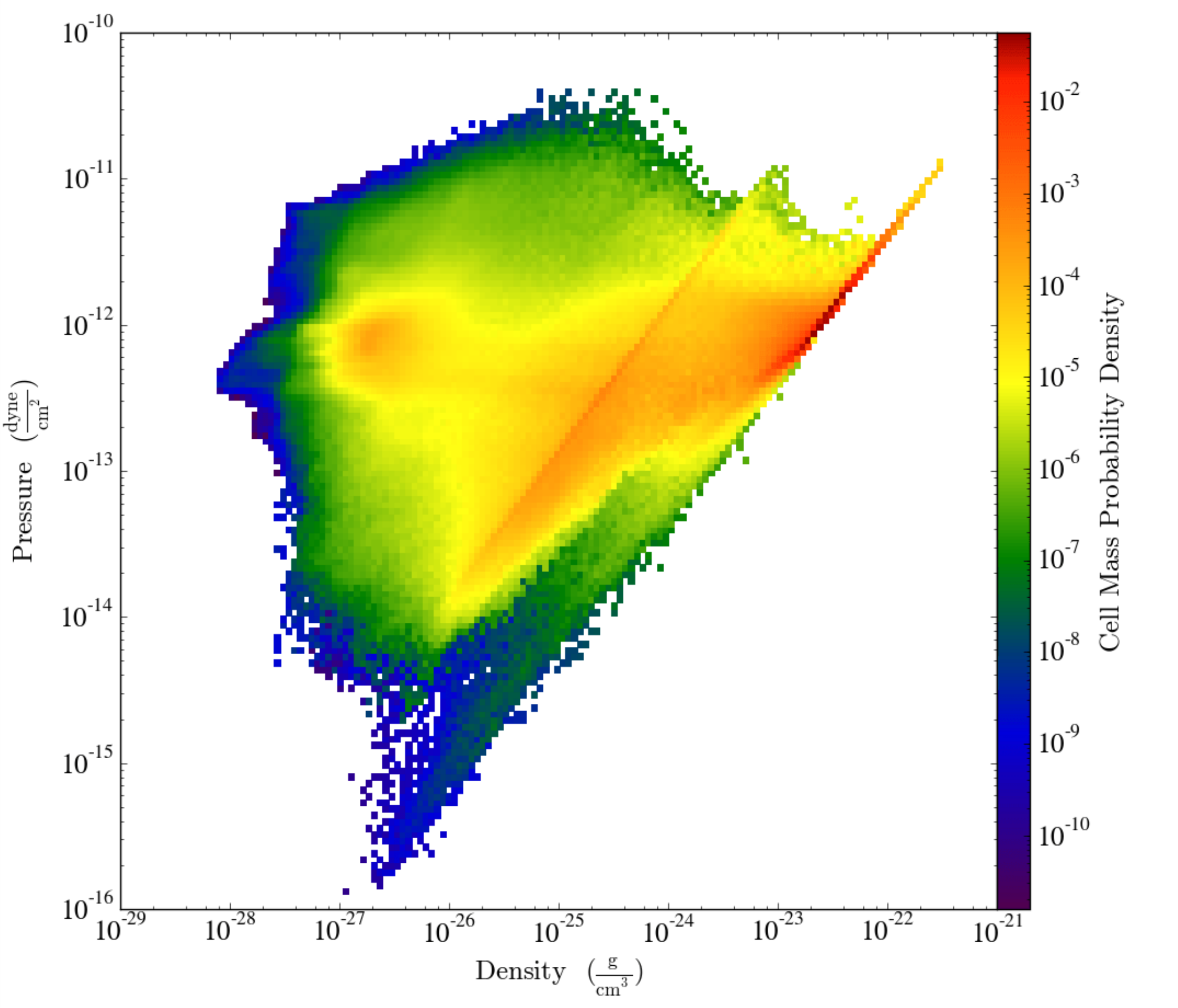}
\caption{Phase diagram of the gas in $\Sigma$10-KS (solar neighborhood model). The color coding shows the fractional mass in each (density, pressure) bin. }
\label{f:phase}
\end{center}
\end{figure} 

Fig \ref{f:phase} shows the phase diagram for the run $\Sigma$10-KS at $t=$ 100 Myr. The color coding indicates the fractional mass in each (density, pressure) bin. The three phases, hot, warm and cold, are clearly seen. Each of the three phases has some spread in the density distribution. But the majority of them are in pressure equilibrium, with $P/k_B \sim 5\times 10^3 \rm{cm^{-3}}$ K. This is in good agreement with the observations near the solar neighbourhood \citep{cox05}. Some mass, which lies in between the two diagonal lines that indicate the standard ``warm" and ``cold" phases, is out of thermal equilibrium \citep{heiles03}.

\begin{figure}
\begin{center}
\includegraphics[width=0.5\textwidth]{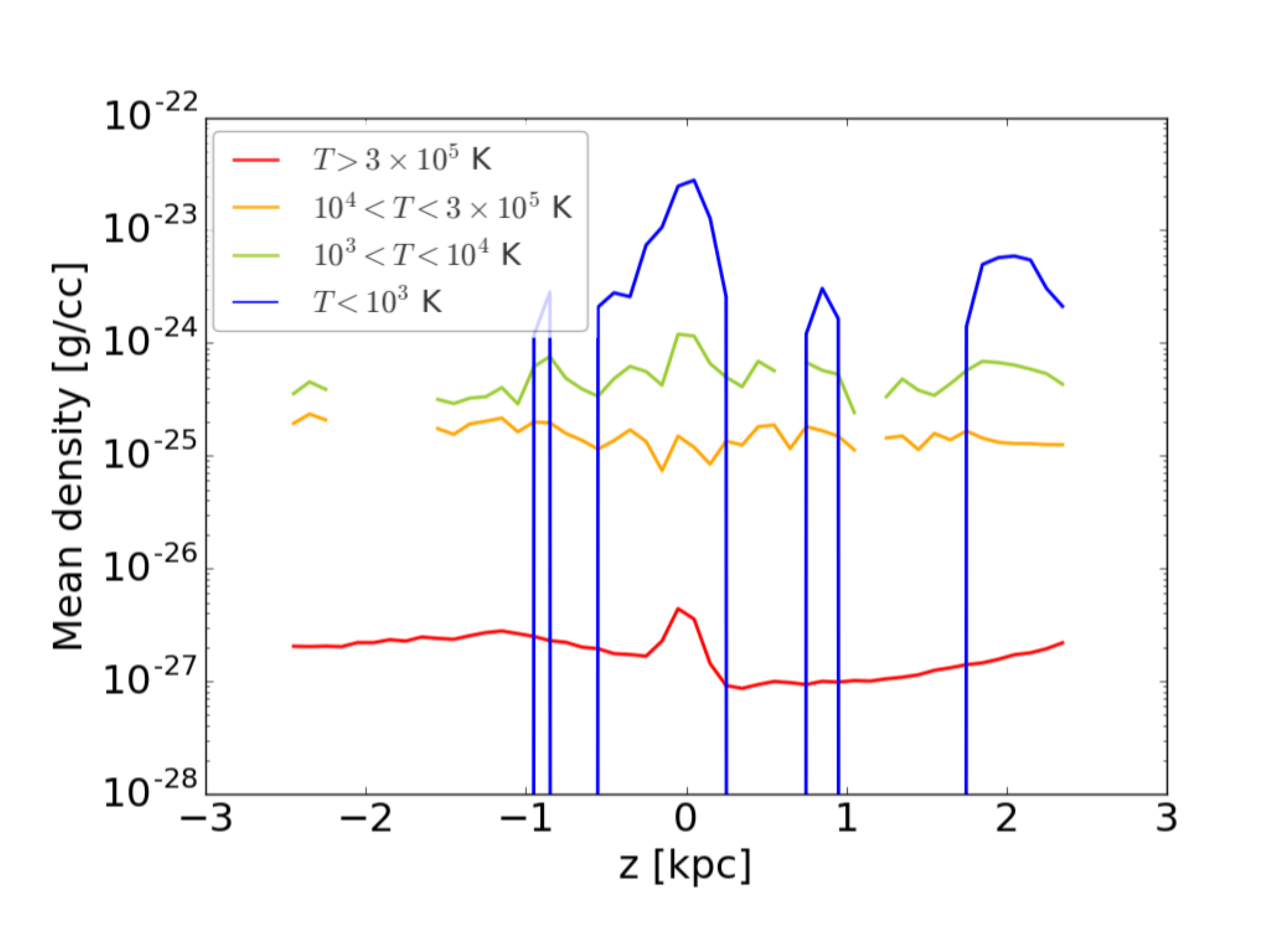} 
\caption{Gas density (volume-weighted) for different phases as a function of z for $\Sigma$10-KS at t$=$ 160 Myr. }
\label{f:nbar}
\end{center}
\end{figure}

Fig \ref{f:nbar} shows the density, weighted by volume, of different phases. 
Hotter phases have progressively smaller densities. Gas density for each phase near the plane is higher than that in the outflows. The warm-hot phase with $10^4 < T < 3\times 10^5$ K has a slightly lower density than the warm phase. As shown in Fig \ref{f:slice_all}, the warm-hot phase is mostly at the interface between warm clouds and hot gas. Even the coldest phase is seen at large $z$, even though the volume fraction can be very small. The densities for  each phase agree with the observations of the local ISM \citep[see, e.g.,][]{draine11}

\subsection{Velocity structure}
\label{sec:vel}
The velocity structure of the gas determines how far the gas can travel in a gravitational potential. Moreover, the velocity distribution can be observed from the profiles of emission/absorption lines. Since different gas phases have drastically different velocities, and observationally, they are detected through different line tracers, we hereby show the velocity distribution for each gas phase, separately.

\begin{figure}
\begin{center}
\includegraphics[width=0.50\textwidth]{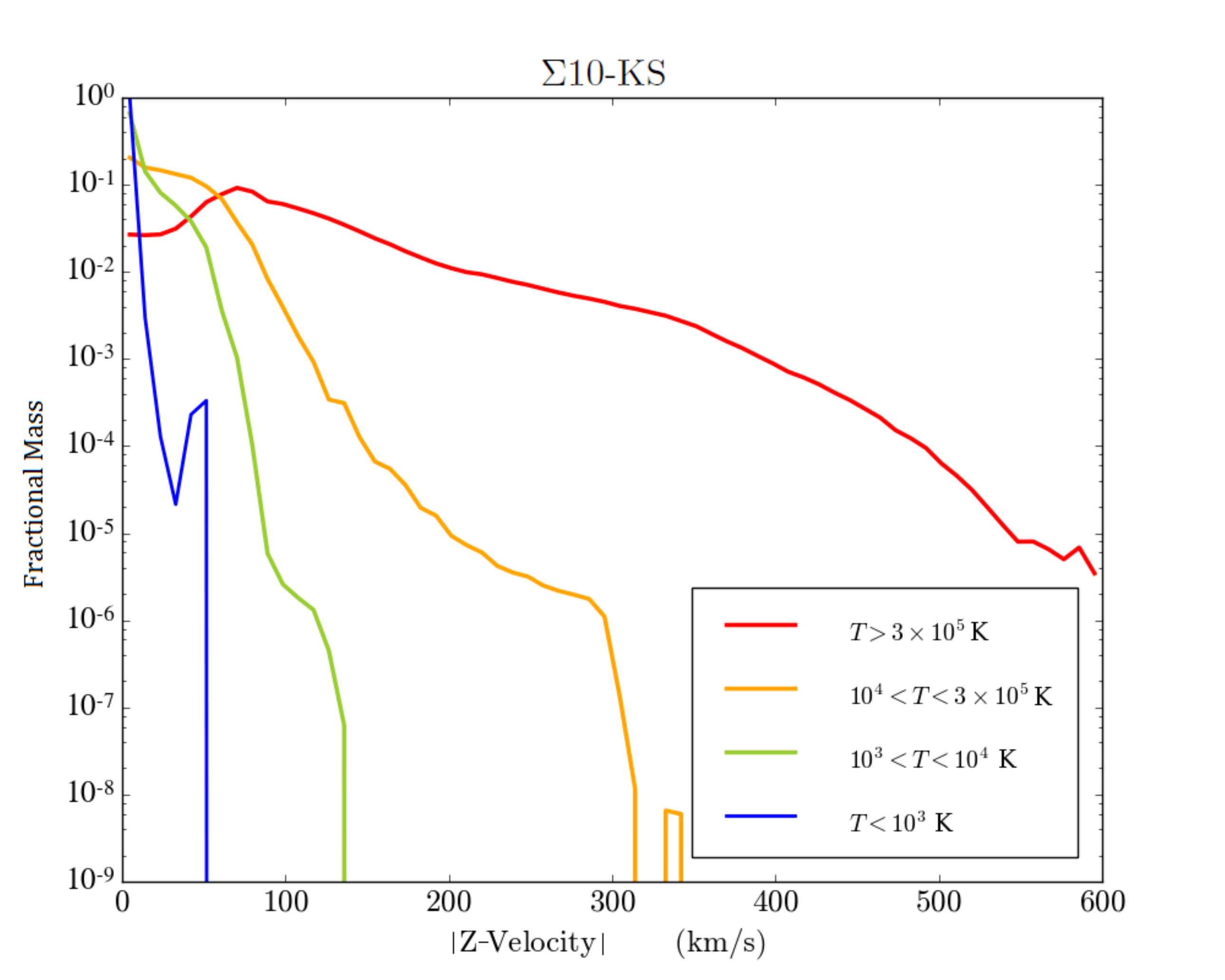}
\caption{Fractional mass for gas with different z-velocity (absolute value),  for the model $\Sigma 10$-KS at $t=160$ Myr. Different curves correspond to gas in different temperature ranges. Each curve is normalized to unity.}
\label{f:z_vel}
\end{center}
\end{figure}

Fig \ref{f:z_vel} shows the z-velocity distribution for the run $\Sigma 10$-KS at $t=160$ Myr. The y-axis indicates the fractional mass in each velocity bin. Each curve is normalized to unity. 
Hotter phases have larger velocities, agreeing with the general observational trend \citep[e.g.][]{heckman01, rupke02}, and other simulation works \citep{creasey13,girichidis16b}. The hottest phase has the broadest range of velocities, up to $\gtrsim$ 600 km/s. A fraction of the warm phase can reach $>$ 100 km/s. The velocities of cold phase remain small at $\lesssim$ 50 km/s.

\begin{figure}
\begin{center}
\includegraphics[width=0.50\textwidth]{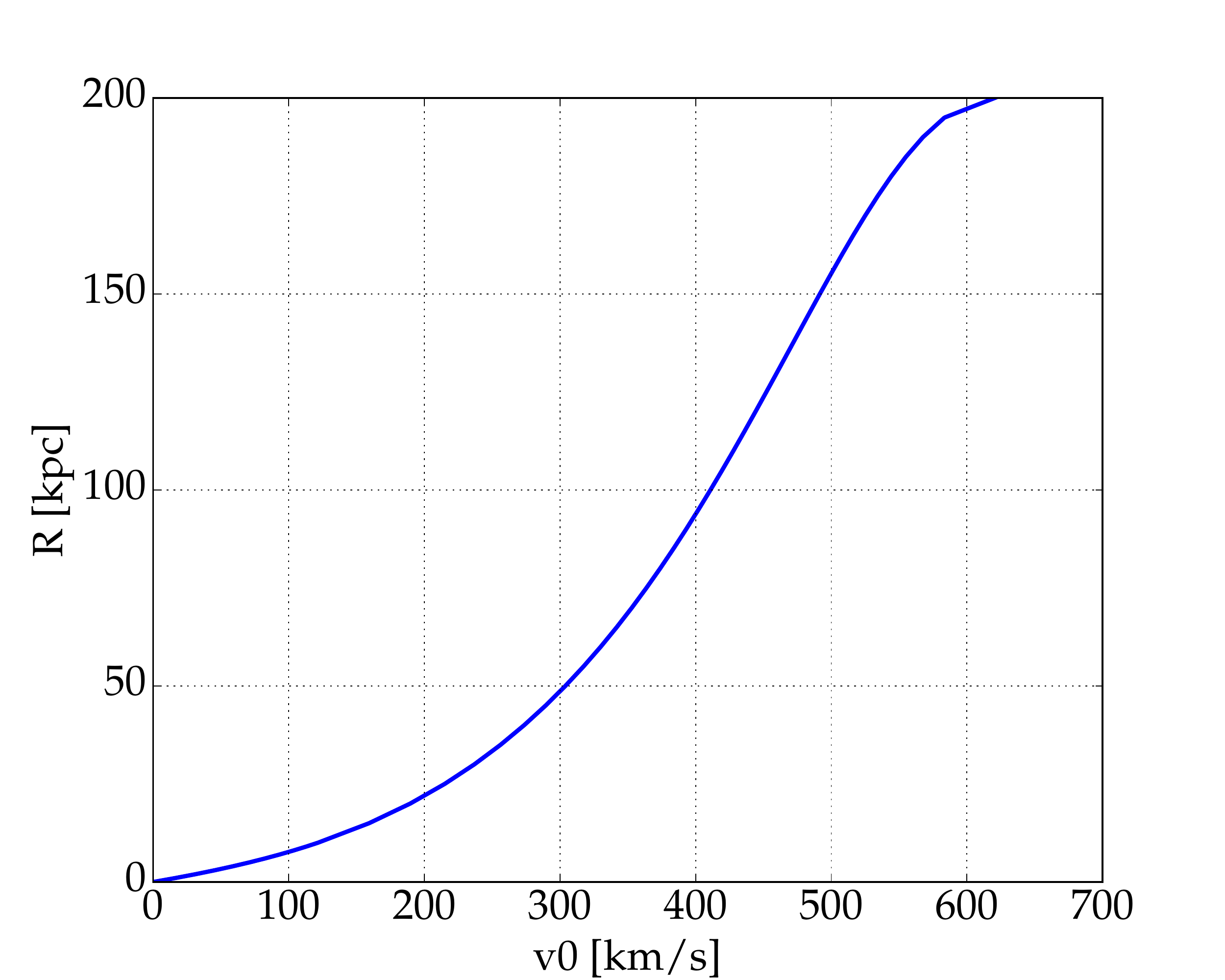}
\caption{Radius $R$ that a parcel of gas with velocity $v_0$ can reach from the center of the MW. See Section \ref{sec:vel} for model details. }
\label{f:v_R}
\end{center}
\end{figure} 

Our simulations only capture the gas evolution that is relatively close to the midplane, i.e. $|z|\leqslant$ 2.5 kpc. One way to relate the ``local'' outflows to their large-scale evolution is to estimate how far the gas can travel in a given potential. First, let us consider a ballistic evolution. For a parcel of gas with a velocity $v_0$ at the bottom of a potential well, the furthest distance it can reach, $R$,  is simply determined by 1/2 $v_0 ^2 = \Delta \phi (R) $.  We use function $R(v_0)$ to describe such a relation. Fig \ref{f:v_R} shows $R(v_0)$ for the MW, for a single stream  line that is perpendicular to the disk and goes through the center of the disk. The potential of the DM halo is the same as described in Section \ref{sec:method}, and the disk is modeled as a 2D razor-thin disk with a mass $M_{\rm{D}} = 5\times 10^5\msun $ and a radius of $R_{\rm{D}}=$ 9.5 kpc. The mass distribution within the disk is uniform.  Thus, along the stream line mentioned above, the g-field from the disk has a simple analytic form: $g_{\rm{D}} =  2 G M_D  (1- z/\sqrt{z^2 +R_{\rm{D}}^2})/R_{\rm{D}}^2$. From Fig \ref{f:v_R}, gas with $v_0 \gtrsim 620$ km/s can escape from the DM halo; gas with $v_0 =$ 300 km/s can travel to  $R\sim$ 50 kpc, and so on. 

We now discuss what should be used as $v_0$. The naive answer, the bulk velocity projected to the direction of $g$, may only give a lower bound. For a compressible fluid, it is likely that the gas motion is not ballistic, but thermal energy can later convert to bulk motions. According to the Bernoulli principle, the Bernoulli constant $\mathcal{B}\equiv v_z^2/2 + \gamma/(\gamma-1) P/\rho + \phi$, remains unchanged along a stream line in a steady-state flow (for a constant $\gamma$).  We thus  define a modified ``Bernoulli velocity'' \vB$\equiv\sqrt{2} \mathcal{\widetilde{B}}^{1/2}$, where $\mathcal{\widetilde{B}} \equiv \mathcal{B} - \phi$. So for a parcel of gas with a bulk velocity $v_z$ and a ``Bernoulli velocity" \vB, the approximate range of radii it can reach is roughly $R(v_0 = v_z)\sim R(v_0 = v_\mathcal{\widetilde{B}}) $. Note we only aim at a very rough estimate, ignoring cooling, interaction among different gas phases, etc, and assuming $\gamma =5/3$.

\begin{figure}
\begin{center}
\includegraphics[width=0.48\textwidth]{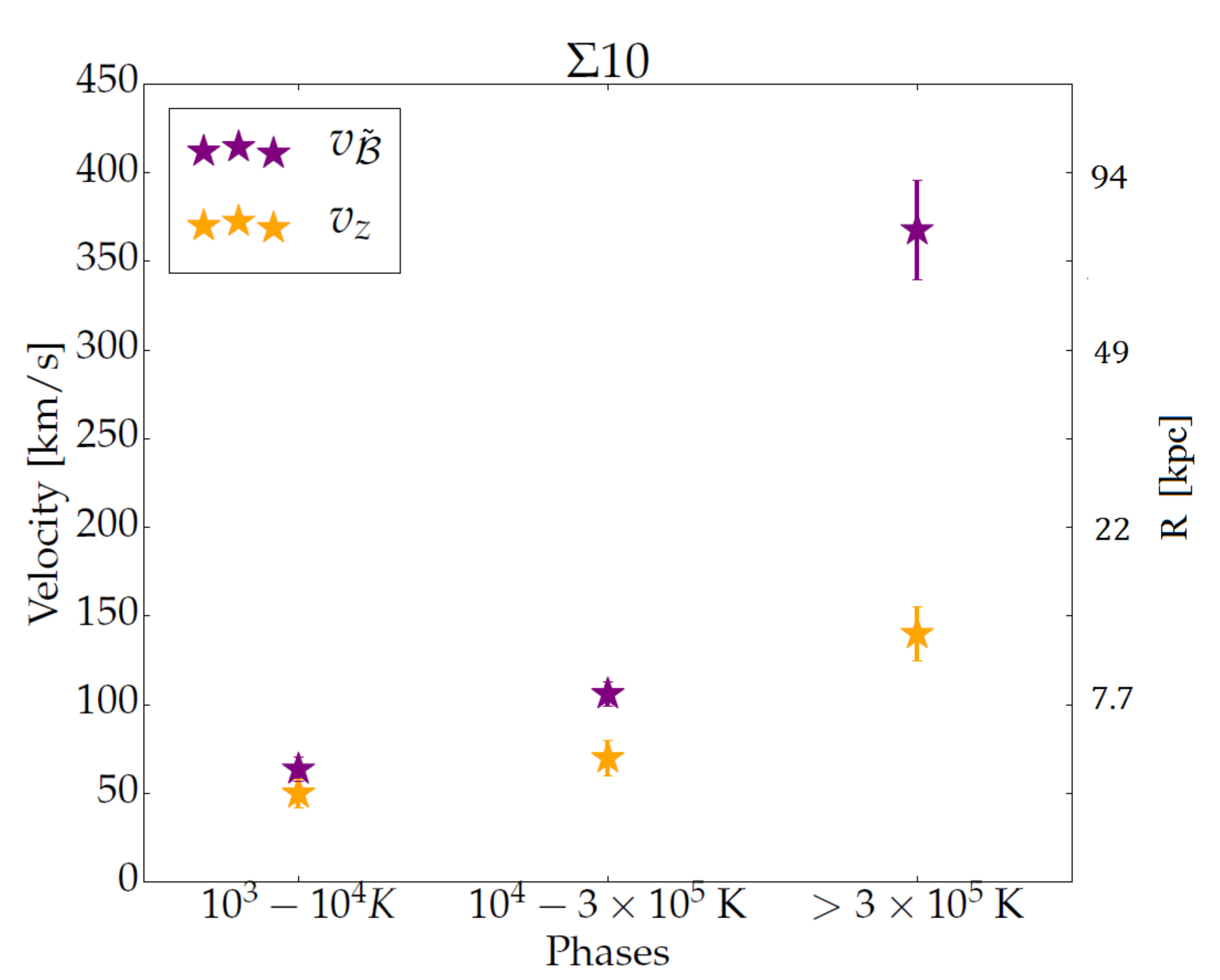}
\caption{Mass-weighted $v_z$ and \vB of the outflows in different temperatures ranges. Y-ticks on the right show $R$ corresponding to the velocities on the left, as in Fig \ref{f:v_R}. See Section \ref{sec:vel} for details.  }
\label{f:MW_vz_vB}
\end{center}
\end{figure}

Fig \ref{f:MW_vz_vB} shows the mass-wighted $v_z$ and \vB for different phases for the fiducial run $\Sigma$10-KS. The y-axis on the right shows $R$ corresponding to the velocities on the left axis. Only gas at $|z|\geqslant$ 1kpc is included. The data are averaged over the last 20\% of the simulation time. The error bars indicate time variations. The hot gas is affected more by each SN explosion, thus its properties vary stronger with time. Both $v_z$ and \vB increase with gas temperature. The hottest phase, given its large $v_z$ ($\sim$ 150 km/s) and \vB ($\sim$ 370 km/s), would travel much further into the halo, to about 30-70 kpc. Since the majority of the hot gas would not escape from the DM halo, large-scale fountain flows would form. The small velocities of the cool phase imply that they would fall back at below 10 kpc, unless being accelerated significantly.  The velocity of the warm-hot phase, with T$=10^4 -3\times 10^5$K, is much closer to the warm phase than the hot. The ratio \vB/$v_z$ is largest for the hot phase, meaning that a significant fraction of the energy is thermal, which may convert to the bulk motion at large radii. For cooler phases, in contrast, most energy is kinetic. Note that the fiducial run is for the solar neighbourhood, and is not representative of the MW disk in general. We discuss the model for the MW-average, $\Sigma10$-KS-4g, in Section \ref{sec:g-field}.  Hot flows are much faster there than the solar neighborhood, and can thus have a much broader impact on the CGM (see Section \ref{sec:g-field} for details).

\begin{figure}
\begin{center}
\includegraphics[width=0.5\textwidth]{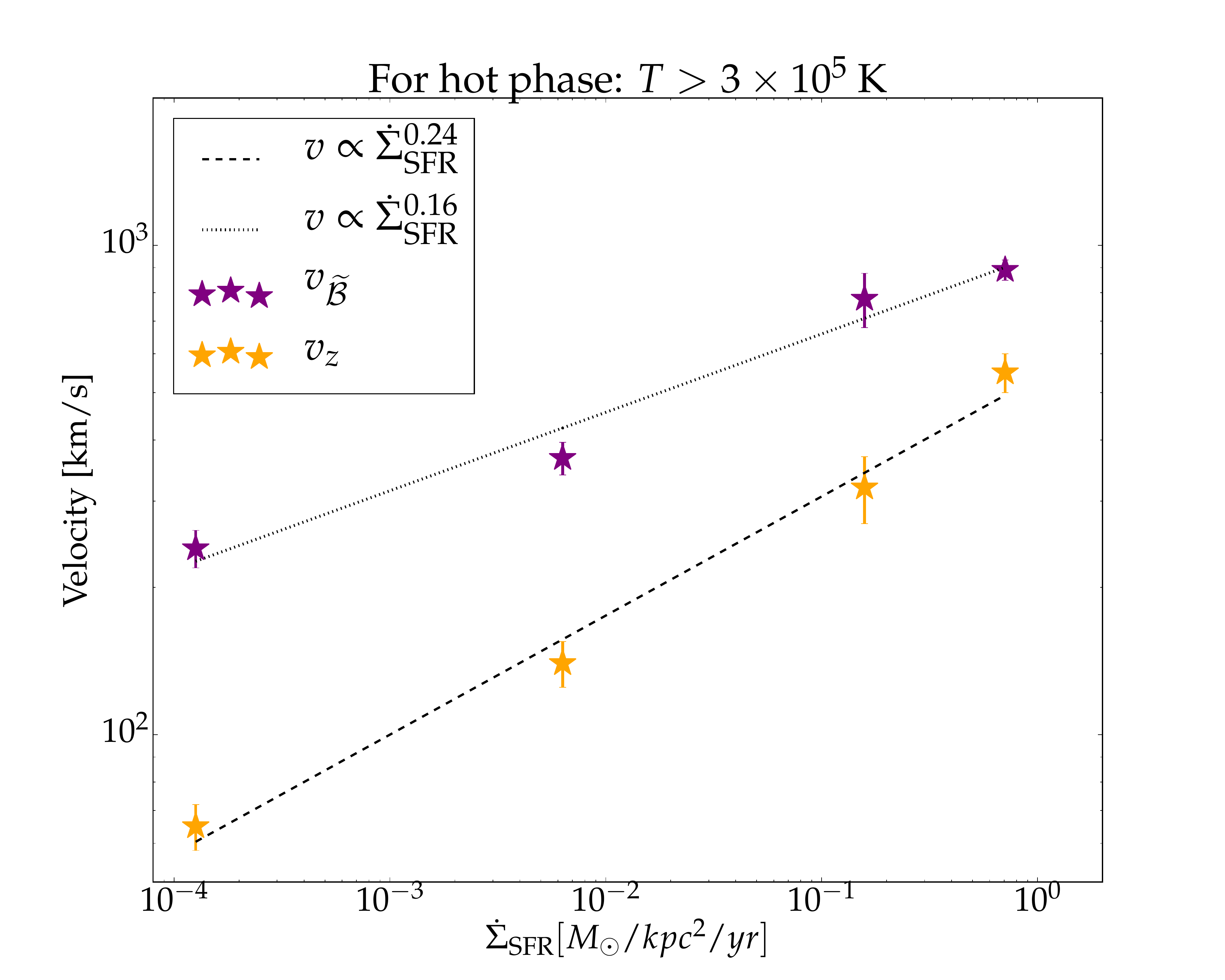}
\caption{Mass-weighted $v_z$ and \vB for hot outflows ($T > 3\times 10^5$ K), as a function of $\dot{\Sigma}_{\rm{SFR}}$, for the four fiducial runs. The power-law fits of the data are indicated in the box. }
\label{f:vz_vB_all}
\end{center}
\end{figure}

Fig \ref{f:vz_vB_all} shows the mass-weighted $v_z$ and \vB for the hot outflows, as a function of $\dot{\Sigma}_{\rm{SFR}}$ for the four fiducial runs. Again, the error bars show time variation. Both $v_z$ and \vB increase with SFR. The velocities for the $\Sigma 1$-KS run are 60-200 km/s, and rise to 600-900 km/s for $\Sigma 150$-KS. The large velocities imply that the hot outflows can travel far, and even escape from the halo potential. This suggests that hot outflows play a critical role on regulating the CGM and even the IGM. 

We find that $v_z \propto \dot{\Sigma}_{\rm{SFR}}^{0.24} $, and $v_{\widetilde{\mathcal{B}}} \propto \dot{\Sigma}_{\rm{SFR}}^{0.16} $. Observationally, while there is little direct constraint on velocities for hot gas, for the warm/cool phases, \cite{martin05} and \cite{weiner09} found $v \propto \rm{SFR}^{0.3-0.35}$ for galactic-scale outflows. Our findings seem to indicate that the dependence of SFR for the hot gas velocity is weaker than for the cooler phases.

\subsection{Loading factors}
\label{sec:load}

In this section, we discuss the loading capability of the outflows. We define outflows to be at $|z|\geqslant 1$ kpc and with outgoing z-velocity. We find in our simulations, the outflow fluxes show little variation with z at $|z|\geqslant 1$ kpc. 

\begin{figure*}
\begin{center}
\includegraphics[width=1.00\textwidth]{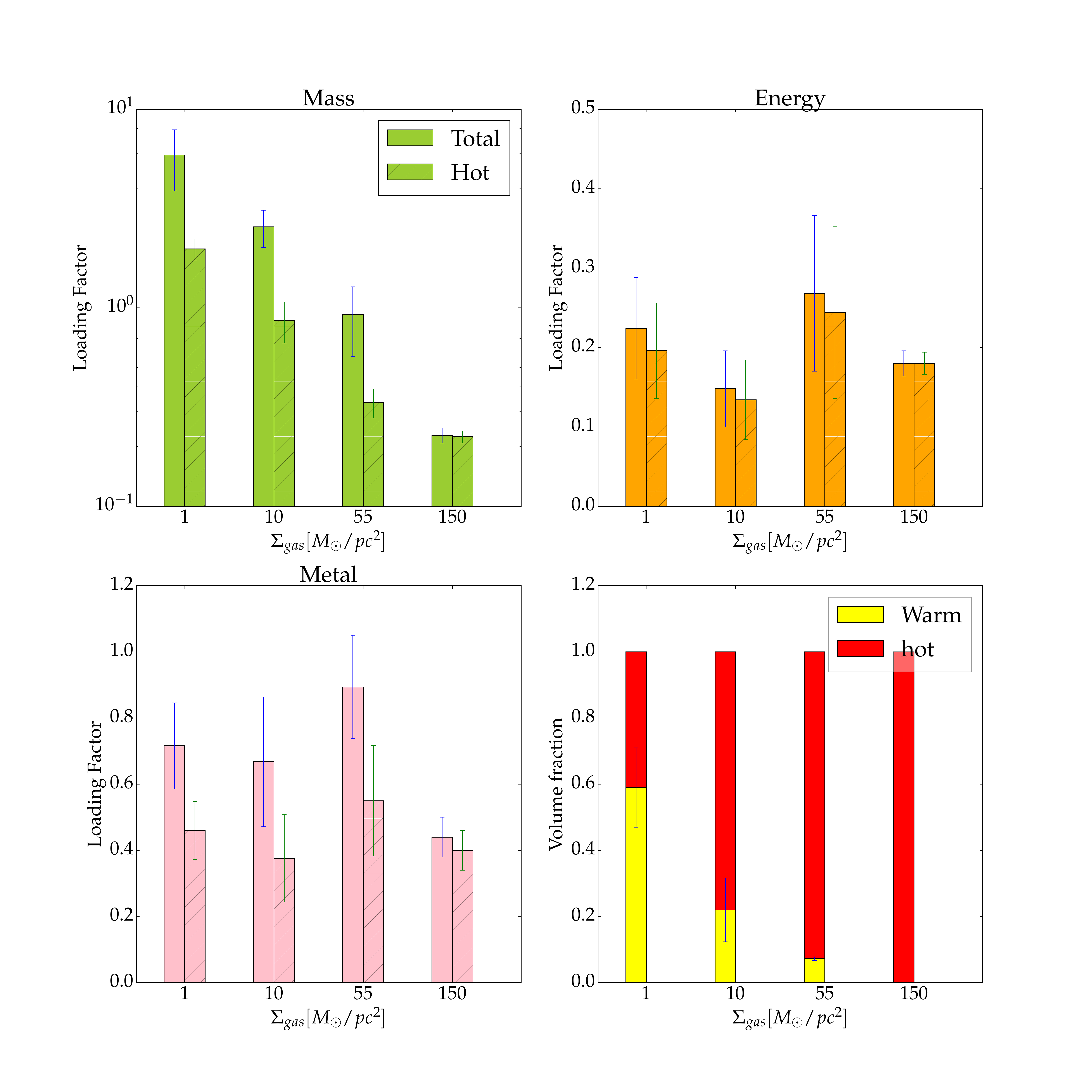}
\caption{Loading factors and volume fraction of each gas phase as a function of $\Sigma_{\rm{gas}}$. The quantities are calculated for outflowing gas at $|z|\geqslant$1 kpc. ``Hot'' and ``warm'' here denote $T>3\times 10^5$ K, and $10^4 < T< 3\times 10^5$ K, respectively. See Section \ref{sec:load} for details. }
\label{f:loading_all}
\end{center}
\end{figure*}

The mass loading factor $\eta_m$ is defined as the ratio between the outflowing mass flux and $\dot{\Sigma}_{\rm{SFR}}$, that is,
\begin{equation}
\eta_{m} \equiv \frac{< \rho v_z >}{ \dot{\Sigma}_{\rm{SFR}}}.
\end{equation}
The outflow flux includes both sides of the plane, and ``$<$...$>$'' denotes averaging over space ($1\leqslant |z| \leqslant  2.5$ kpc) and time (last 40\% of $t_{\rm{sim}}$).

The energy loading factor $\eta_E$ is the ratio between the z-component of the energy flux and the energy production rate by SNe, that is, 
\begin{equation}
\eta_{E} \equiv  \frac{< (e_k + e_{th}) v_z >}{\dot{\Sigma}_{\rm{SFR}} E_{\rm{SN}}/ m_0},
\end{equation}
where $e_k$ and $e_{th}$ are the kinetic and thermal energy per unit volume. 

The metal loading factor $\eta_{met}$ is the ratio of the z-component of the metal flux to the metal production rate by SNe.
\begin{equation}
\eta_{\rm{met}} \equiv \frac{<\rho_{\rm{met}}v_z> }{\dot{\Sigma}_{\rm{SFR}}\  m_{\rm{0,met}}/m_0},
\end{equation}
where $\rho_{\rm{met}}$ is the density of metals, and $m_{\rm{0,met}}$ is the mass of metals each SN produces (in arbitrary units, see Section \ref{sec:SN_model} ). Note that we assume the metals are solely produced by SNe, and the ISM is otherwise pristine. 

Fig \ref{f:loading_all} summarizes the loading factors and the volume fraction of each gas phase in the outflows as a function of $\Sigma_{\rm{gas}}$, for the four fiducial runs. For the loading factors we show the total loading, which includes all gas phases, as well as that of the hot outflows only. The error bars indicate the standard deviation of the time variation.

We find that \ml decreases monotonically with increasing $\Sigma_{\rm{gas}}$. The largest mass loading is about 6 for $\Sigma$1-KS. For the solar neighborhood case, i.e. $\Sigma$10-KS, our \ml is around 2-3. For our highest density case $\Sigma$150-KS, \ml is only 0.2. The fraction of the mass loading contributed from the hot gas is about 1/3, except for $\Sigma150$-KS, where most of the mass flux is hot. The warm phase dominates the outflowing mass flux except when \sigSFR is very high. 

We fit our mass loading factor by a simple power-law function of \siggas:
\begin{equation}
\eta_m = 7.4 (\frac{\Sigma_{\rm{gas}}}{1 \msun/\pc^2})^{\alpha_{\rm{ml}}}, \ \alpha_{\rm{ml}} = -0.61 \pm 0.03.
\label{eq:ml}
\end{equation}
For the hot gas, the mass loading factor
\begin{equation}
\eta_{m,h} = 2.1 (\frac{\Sigma_{\rm{gas}}}{1 \msun/\pc^2})^{\alpha_{\rm{ml,h}}}, \ \alpha_{\rm{ml,h}} = -0.61\pm 0.03.
\end{equation}
\cite{creasey13} have found a sharper decline, with $\alpha\approx$ -1.1. Our results agree with theirs for \siggas$\lesssim 10\msun/\pc^2$, but there are relatively large discrepancies at higher densities. See Section \ref{sec:compare} for a discussion. The X-ray emission from the halo of edge-on galaxies suggests a decreasing mass loading of hot gas for higher SFR \citep{zhang14, bustard16}, consistent with our results. 

The energy loading factor \el shows surprisingly little dependence of \siggas. Despite a factor of 150 span of \siggas, \el stays at about 10-30\%. This means a significant fraction of SNe energy goes into the outflows.  There is no obvious trend of \el as a function of \siggas. The hot gas contains the majority, $>$90\%, of the outflow energy.  

The metal loading factor \mel shows somewhat larger variation than \el, although again we do not find an apparent dependence on \siggas. Overall, a quite large fraction of metals go into the outflows, about 40-90\%. Hot outflows carry 35-60\% of the metals produced by SNe. While the warm/cool phase may fall back to the disk later, the hot gas has the potential to travel much further, even escape the halo (see Section \ref{sec:vel}), and metals will be carried along. The mass-metallicity relation of galaxies implies that a significant fraction of metals ever produced are no longer in galaxies \citep{tremonti04,erb06}. Our numbers agree with this general picture. 

The volume in outflows is progressively occupied by the hot gas as \siggas becomes larger. The cold phase with $T<10^3$ K has a negligible volume fraction, thus we omit it in the plot. For $\Sigma$1-KS, the volume is equally shared by warm and hot phase; for $\Sigma$55-KS, more than 90\% of volume is hot; for $\Sigma$150-KS, the outflows are completely dominated by the hot phase.

\section{Effects of several physical processes}
\label{sec:multi-effect}

\subsection{SNe scale height}
\label{sec:SN_height}

Where SNe explode is critical for feedback efficiency. A SN exploding in a dense medium quickly radiates away its energy, and has little impact on the large-scale ISM, let alone contributing to driving winds \citep{girichidis16b}. On the other hand, if a SN explodes in an environment dominated by tenuous gas, then the cooling is much less efficient, and a significant fraction of energy can be preserved \citep[e.g.][]{li15, gatto15,simpson14,walch15, hennebelle14}. One key factor to determine where SNe explode is the fact that a significant fraction of OB stars are ``runaways", that is, having high velocities. A simple calculation shows that OB runaways can migrate a few tens to a few hundred pc before exploding as SNe \citep{li15}. This greatly facilitates SN feedback by allowing some of them to release their energy outside the dense SF regions. 

In principle, the locations of core collapse SNe depend on the velocities of OB stars, their lifetimes, the external gravitational field, close encounters with other stars, etc. One can also infer the SNe explosion sites from the spatial distribution and the velocities of pulsars \citep{narayan90}. In this paper, we do not aim to model the location of SNe from first principles, but simply explore how sensitively the outflow properties depend on the vertical distribution of SNe. For the fiducial runs we have the $h_{\rm{SN,cc}}=$150 pc. Now we experiment with  $h_{\rm{SN,cc}}$. We take the run of $\Sigma$55-KS and change \hSN to 75, 300, 450 pc, respectively. In Table 1, they are identified by names of $\Sigma55$-KS-h75, $\Sigma55$-KS-h300, and so on.

\begin{figure}
\begin{center}
\includegraphics[width=0.50\textwidth]{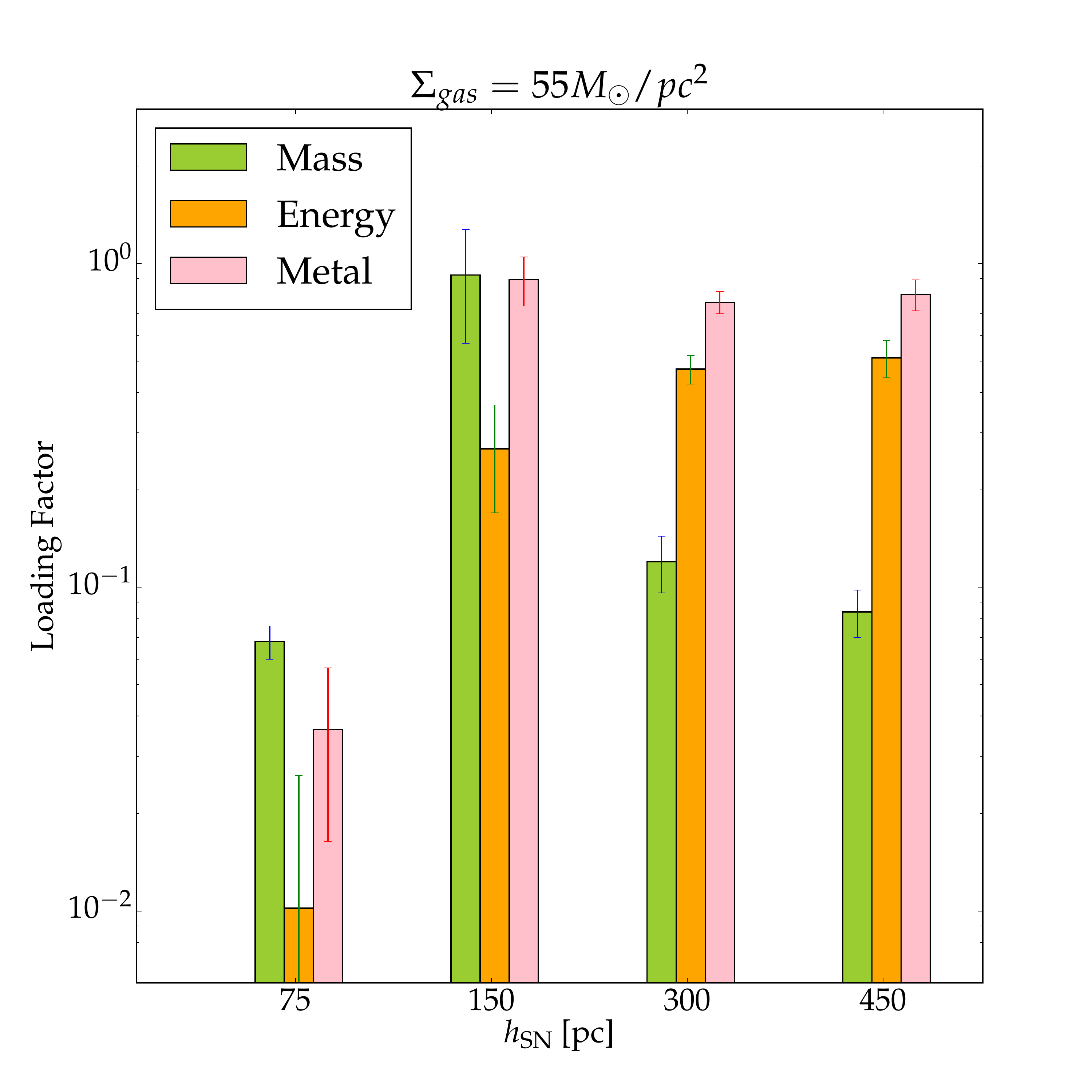}
\caption{Loading factors for different SN scale heights \hSN\ for the model $\Sigma 55$-KS. See Section \ref{sec:SN_height} for details.  }
\label{f:h_SN_compare}
\end{center}
\end{figure}

Fig \ref{f:h_SN_compare} shows the mass, energy and metal loading factors for different \hSN. Interestingly, \ml depends on \hSN\ in a different way from \el and \mel. As \hSN\ increases, \ml first increases, and reaches the peak at \hSN$=150$ pc, and then declines. Loading factors \el and \mel increase monotonically with \hSN, and reaches plateau at \hSN$\gtrsim$150-300 pc. For \hSN$=$75 pc, most SNe are buried in the mid-plane, and radiate their energy there, so the feedback is least efficient. As \hSN\ becomes larger, more SNe explode in the low density disk-halo interface, resulting in a more effective energy and metal loading. The non-monotonic dependence on \hSN\  of \ml can be understood in this way: when \hSN\ is too small, most energy radiates away in the disk, so the energy insufficiency is the limiting factor for the mass loading; when \hSN\ is too large, the outflowing mass is simply SNe ejecta, with little ISM involved. Consequently, the maximum \ml occurs in between those two extremes. We find that \ml achieves unity when \hSN $=$150 pc, while \ml $\lesssim $ 0.2 for other cases. For \hSN $\gtrsim$ 300 pc, 40-50\% of energy and 80\% of metals produced by SNe end up in the outflows.

\subsection{Photoelectric heating}
\label{sec:PEH}

In the absence of SN explosions, PEH maintains a two-phase warm/cold ISM \citep{draine78, wolfire95}. The value of the PEH rate $\Gamma_{\rm{PEH}}$ determines the relative amount of mass in the two phases and the pressure of the ISM \citep{wolfire03}. With SNe,  $\Gamma_{\rm{PEH}}$ is an important factor in determining whether the ISM is in a thermal runaway state or not \citep{li15}. A higher $\Gamma_{\rm{PEH}}$ keeps more gas in the warm phase and increases the ISM pressure, thus limiting the size of the hot bubbles of SNRs. As a result, SNRs may not effectively overlap, and each SNR loses the majority of its energy at the cooling stage. Therefore little energy is left to drive an outflow. We thus expect that  $\Gamma_{\rm{PEH}}$ is important in determining the outflow properties.

In this section we study the effect of different values of $\Gamma_{\rm{PEH}}$. We note that  $\Gamma_{\rm{PEH}}$ depends on many factors: the far-UV background, dust abundance, work function of the dust grains, ionization fraction of the gas, etc \citep{draine11}. A star forming region has a very complex structure with strong and time-varying radiation background with both ionizing and non-ionizing photons. Radiation background also varies in space, and is much more intense around OB stars. The exact condition is thus hard to determine. For simplicity, we keep $\Gamma_{\rm{PEH}}$ constant in time and uniform in space for each simulation, but just change  $\Gamma_{\rm{PEH}}$ to explore its effect. Note that some previous works adopt a cooling curve with a cut-off at $10^4$ K, which prohibits the formation of the cold phase. This is similar to the effect of a very high  $\Gamma_{\rm{PEH}}$. We include a discussion of the cooling curve cut-off  as well. 

We compare four runs that have \siggas $=55\msun /\pc^2$. The set-ups are identical (including the SN rate) except $\Gamma_{\rm{PEH}}$: \\
(a) $\Gamma_{\rm{PEH}} = 0$ (``$\Sigma$55-KS-noPEH"); \\
(b) $\Gamma_{\rm{PEH}} = 3.5 \times 10^{-25}$ erg/s (fiducial, ``$\Sigma$55-KS"); \\
(c) $\Gamma_{\rm{PEH}} = 1.75 \times 10^{-24}$ erg/s (``$\Sigma$55-KS-5PEH");\\ (d) cooling curve has a cut-off at $T_{\rm{min}}=10^4$ K (``$\Sigma$55-KS-1e4K"). \\

\begin{figure}
\begin{center}
\includegraphics[width=0.5\textwidth]{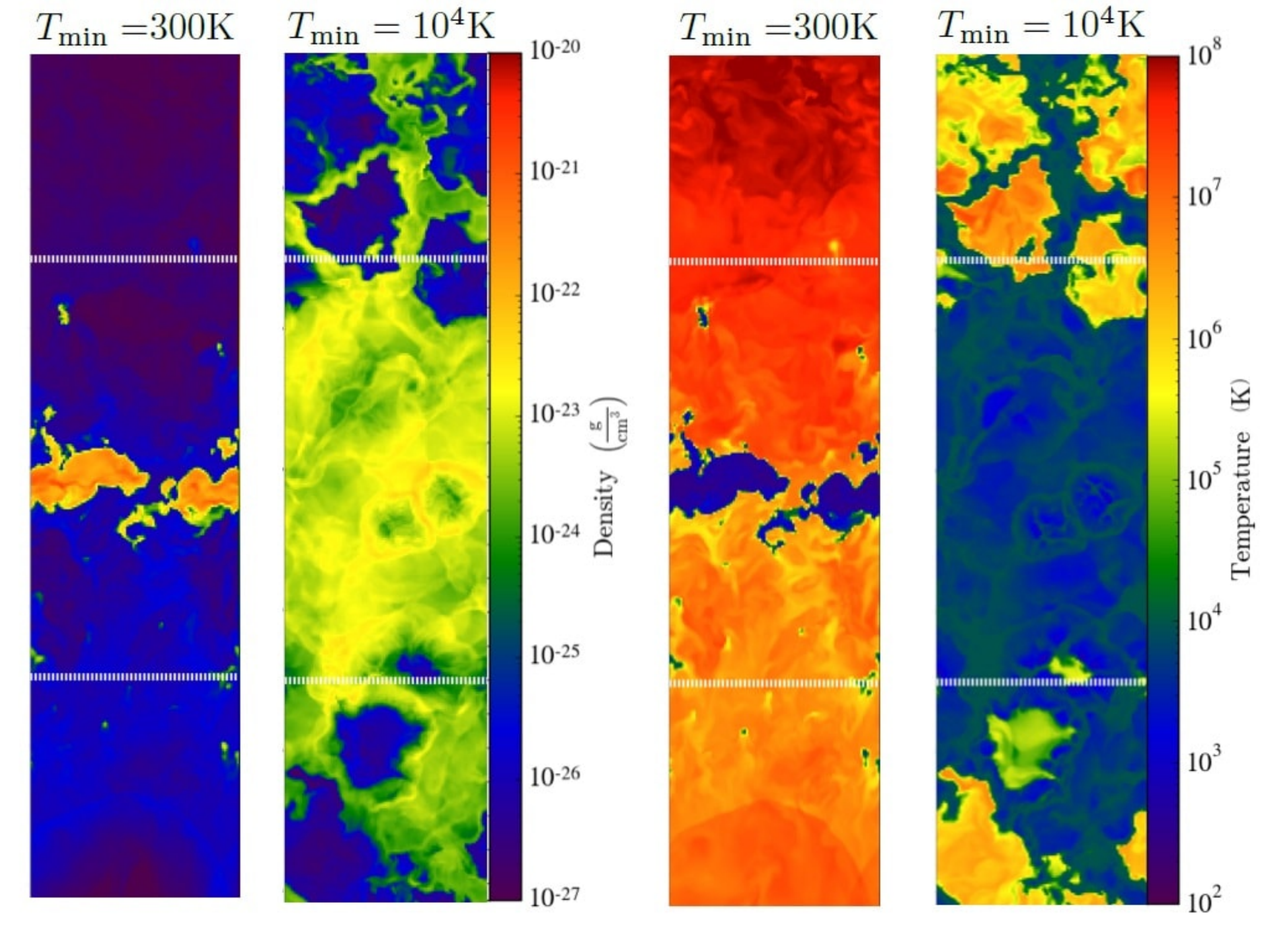}
\caption{Temperature and density slices of two runs with \siggas$=$55 $\msun/\pc^2$: left - fiducial ($\Sigma$55-KS); right - cooling curve has a cut-off at $T_{\rm{min}}= 10^4$ K  ($\Sigma$55-KS-1e4K). The snapshots are taken at $t= 41$ Myr. The slices only include region at $|z|\leqslant 325$ pc. The white dashed lines indicate the scale height of core collapse SNe $h_{\rm{SN,cc}} =$150 pc. Cut-off of the cooling curve at $10^4$ K results in a much larger gas scale height. As a result, most SNe energy is lost through radiative cooling in the dense gas layer. } 
\label{f:300_1e4K}
\end{center}
\end{figure}

In Fig \ref{f:300_1e4K} we show the slices for the fiducial run and $\Sigma$55-KS-1e4K. Adopting $T_{\rm{min}}=10^4$ K results in a much larger scale height of gas (defined as enclosing 80\% of the mass in the box), $h_{\rm{gas}} \sim$ 200 pc, in contrast to $h_{\rm{gas}} \sim 10$ pc for the fiducial case. Assuming hydrostatic equilibrium, i.e., gravity is balanced by the thermal and turbulence pressure,
\begin{equation}
h_{\rm{gas}} \sim 140 \rm{pc}\ (\frac{1+ \mathcal{M}^2}{2}) (\frac{T_{\rm{gas}}}{10^4 \rm{K}}) (\frac{5 \times 10^{-9} \rm{cm/s^2} }{g}),
\label{eq:hgas}
\end{equation}
where $\mathcal{M}$ is the local Mach number of the gas, which is on the order of unity. Note that in our simulations, SNe have a Gaussian distribution with \hSN $=$150 pc. So for the fiducial case, once the multiphase medium is formed, most SNe explode \textit{outside} of the gas layer, whereas for $\Sigma$55-KS-1e4K, most SNe explode \textit{within} the gas layer. For the latter, since the ISM is not in a thermal runaway state, most energy released from SNe is radiated away. Therefore, a cooling curve with $T_{\rm{min}}= 10^4$ K gives much smaller energy, mass and metal loading. It is true that we are adopting a temperature cut of 300 K, and the actual temperature of the cold phase can be even lower. But our temperature cut is low enough to allow the ISM at midplane to undergo a thermal runaway -- as mentioned in Section \ref{sec:multiphase}, all fiducial runs have a volume fraction of hot gas of 60-80\% at midplane. We thus believe that our results do not suffer from a qualitatively erroneous cooling loss, while a temperature cut at 10$^4$ K may do so.

\begin{figure}
\begin{center}
\includegraphics[width=0.48\textwidth]{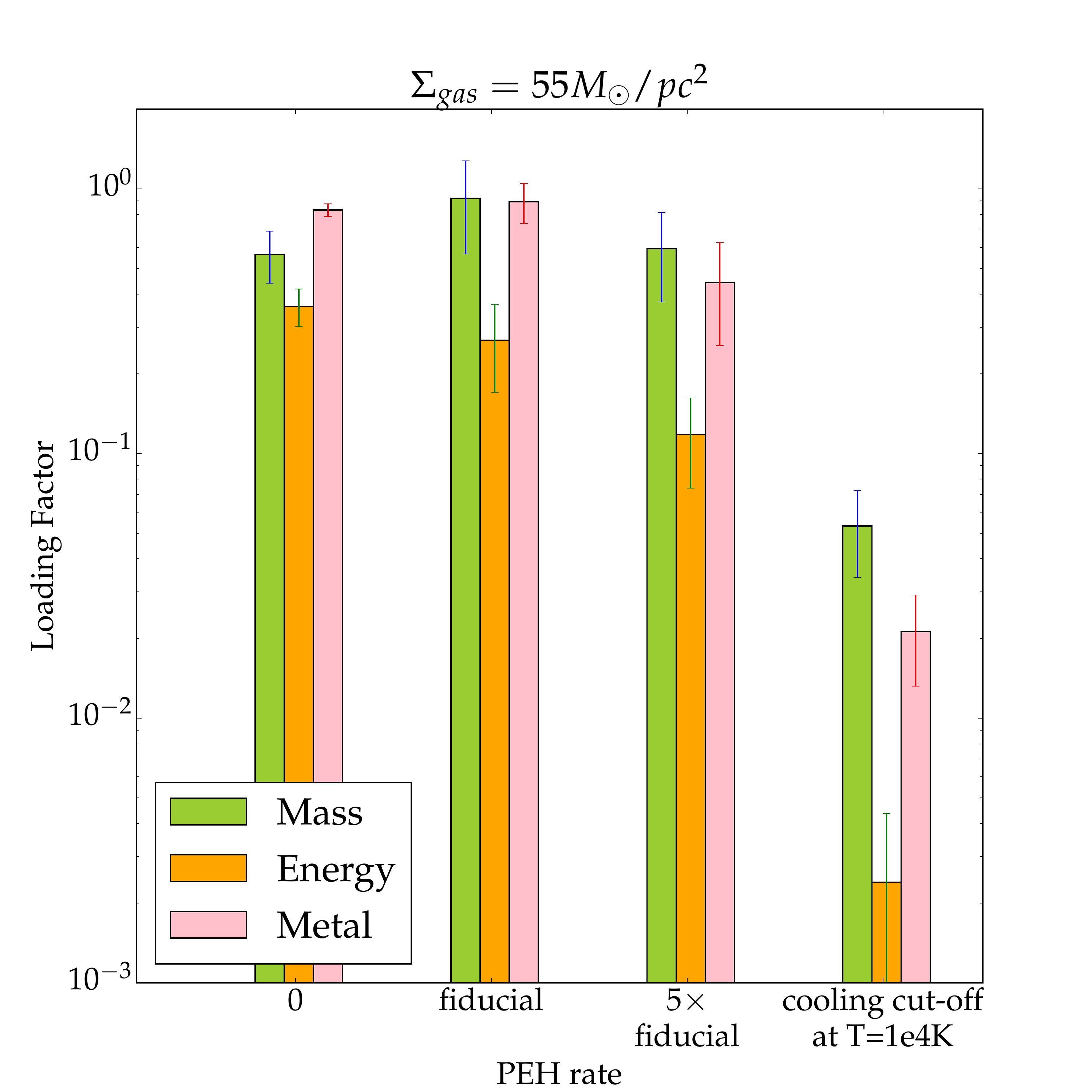}
\caption{Loading factors for different PEH rates, for the model with \siggas $=55 \msun/\rm{pc}^2$. See Section \ref{sec:PEH} for details.  }
\label{f:loading_300_1e4K}
\end{center}
\end{figure}

Fig \ref{f:loading_300_1e4K} compares the loading factors of all four runs in this section. The simulation $\Sigma$55-KS-1e4K gives an energy loading two orders of magnitude smaller than the fiducial run; \ml is smaller by a factor of 10, and \mel by a factor of 30. This indicates that if the formation of the cold phase is prohibited, the power of SN feedback is severely underestimated. Comparing the three runs with different $\Gamma_{\rm{PEH}}$, we find that when $\Gamma_{\rm{PEH}}$ is higher, the energy loading is smaller, as expected, since \hgas\ is increasingly larger for stronger PEH. The mass and metal loadings do not show significant variation. This is likely due to the following two effects counteracting each other: a smaller \el means that less energy is available to drive the mass out, while a larger \hgas\ is favourable to loading more gas, as discussed in Section \ref{sec:SN_height}.

\subsection{External gravitational field}
\label{sec:g-field}

\begin{figure}
\begin{center}
\includegraphics[width=0.5\textwidth]{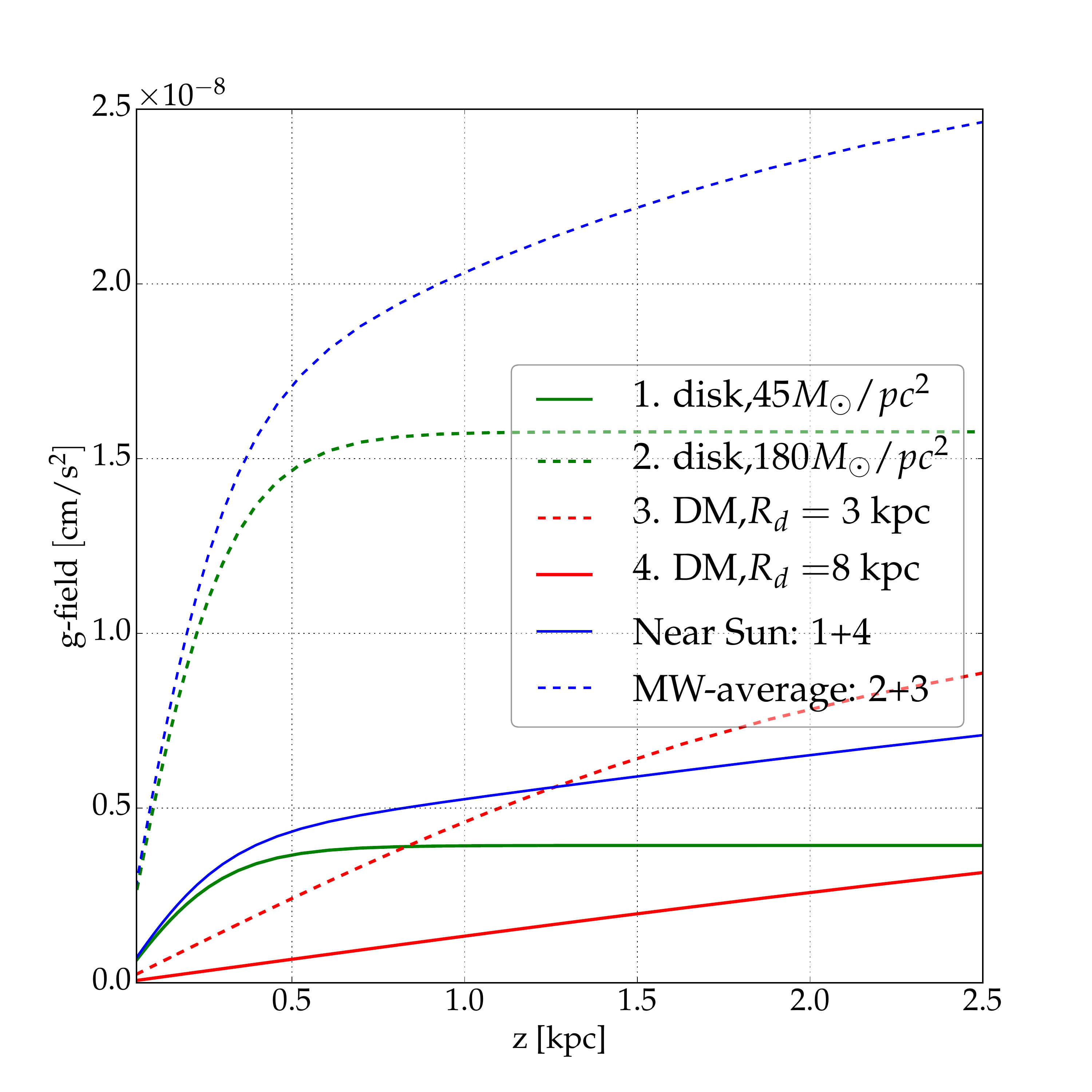}
\caption{External gravitational fields adopted for the solar neighbourhood ($\Sigma10$-KS) and the MW-average ($\Sigma10$-KS-4g). Different components are shown separately. See Section \ref{sec:g-field} for relevant discussions. }
\label{f:g_solar_vs_MW_ave}
\end{center}
\end{figure}

To explore the effect of external gravitational field on the outflows, we take the MW for an example. The fiducial run $\Sigma 10$-KS uses the gravitational field in the solar neighbourhood, which has \siggas $= 10\msun /\pc^2$, $\Sigma_* = 35 \msun /\pc^2$, and a displacement $R_d = 8$ kpc from the center of DM halo. This g-field is smaller than the inner part of the MW disk. We set up a higher gravity run $\Sigma$10-KS-4g, which uses a g-field more typical for the inner MW disk, with \siggas$ = 10\msun /\pc^2$,  $\Sigma_* = 180 \msun /\pc^2$ and $R_d = 3$ kpc. The g-field is approximately 4 times that of the solar neighbourhood. Fig \ref{f:g_solar_vs_MW_ave} shows the different components of the two g-fields.

\begin{figure}
\begin{center}
\includegraphics[width=0.5\textwidth]{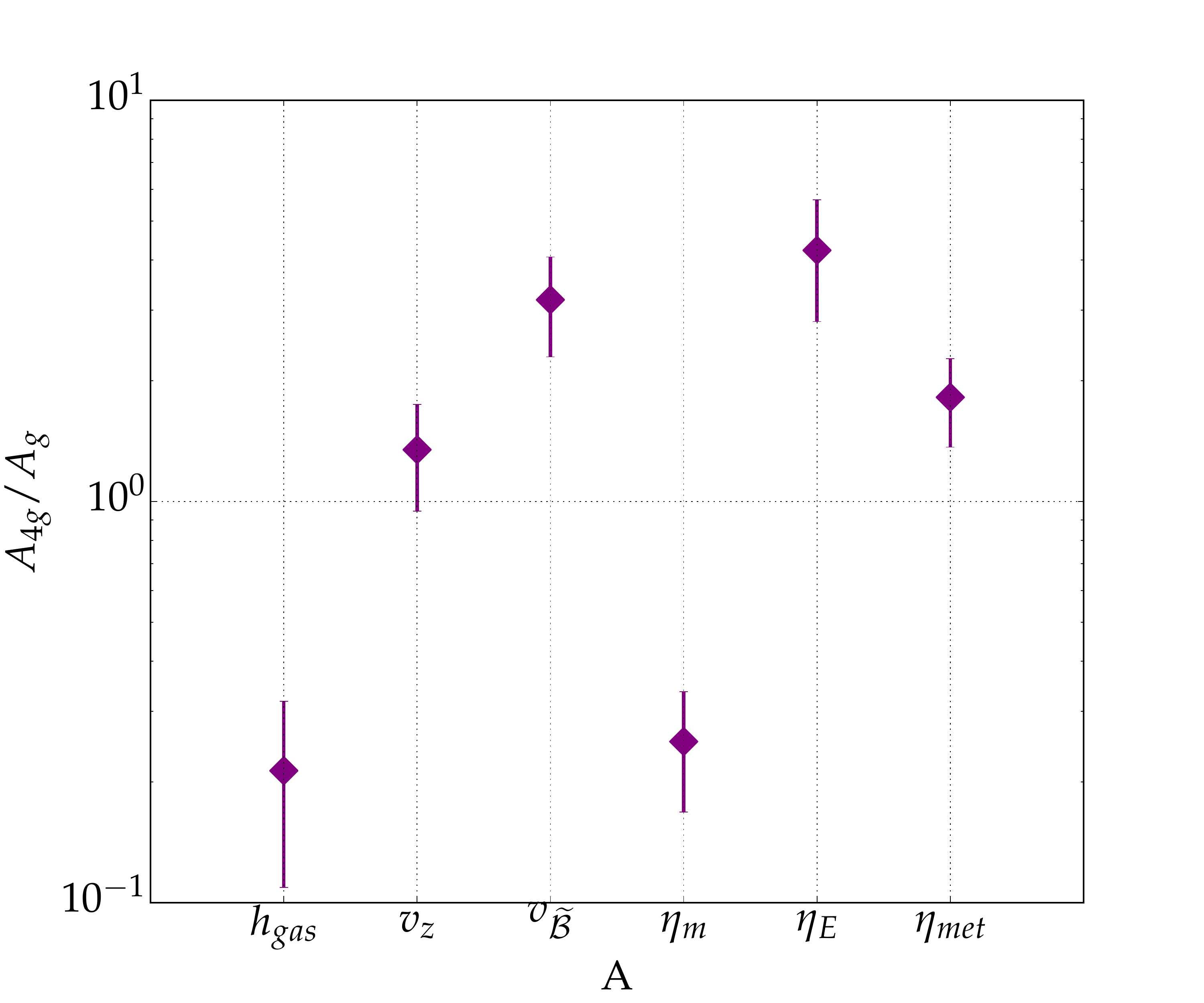}
\caption{Ratio of properties between two runs with different gravitational fields: $\Sigma$10-KS-4g (MW-average) and $\Sigma 10$-KS (solar neighbourhood). See Section \ref{sec:g-field} for the discussion.  }
\label{f:g_field}
\end{center}
\end{figure}

Naively, one would think a larger gravity would make the feedback less effective, as gravity drags the outflows toward the disk. While this is generally true for regions away from the launching region, the situation near the disk is more complex. We plot the ratios of the loading factors between $\Sigma 10$-KS-4g and  $\Sigma 10$-KS in Fig \ref{f:g_field}. The comparison is done for the time interval $t=40-64$ Myr, and the error bar shows the standard deviation of time variation. While \ml is indeed smaller by a factor of 3-4 in the higher gravity case, the energy loading \el is, nevertheless, a factor of 3-5 larger. The metal loading \mel is also larger by a factor of 1.5. 

How to understand this? It turns out that the dominant impact of larger g-field here is to reduce \hgas. As we also show in Fig \ref{f:g_field}, a factor of 4 increase in gravity results in roughly the same factor of decrease in $h_{\rm{gas}}$. This is expected for gas in hydrostatic equilibrium (Eq. \ref{eq:hgas}). Since we keep \hSN\ the same for the two runs, a smaller \hgas\ exposes more SNe in the low-density halo. As discussed in Section \ref{sec:SN_height}, more SNe above the gas layer can lead to a smaller \ml while larger \el and \mel. Since less mass is heated by more energy, \vB of the outflowing gas is much larger in $\Sigma$10-KS-4g, by a factor of 3 than the solar neighbourhood. The values of $v_z$ and \vB are about 175 km/s and 980 km/s, respectively, the latter is even larger than the escape velocity of the MW halo $\sim 620\ \rm{km/s}$. Thus the outflows for the MW-average is much more vigorous, which can broadly impact the CGM and even the IGM (see more discussion in Section \ref{sec:CGM}).

\subsection{Enhanced SNe rates}
\label{sec:SN_rate}

For our fiducial runs, we assume that \sigSFR\ scales with \siggas as in the KS relation. Although the KS relation is well-established on scales $\gtrsim $ kpc,  variation appears on smaller scales \citep{heiderman10}.  In particular, star formation tends to occur in groups, and the OB stars are clustered in space and time. The size of our simulation boxes are in the sub-kpc regime, so it will be interesting and relevant to discuss the variation on the SN rate.  In this section, we discuss the effect of enhanced SN rates on the outflows. We are interested in how the energy, mass, and metal loading efficiencies depend on the SN rates. Since the interaction of blast waves is highly non-linear, it is non-trivial to predict whether the impact of multiple SNRs would be a simple add-up, or to reinforce, or to cancel out each other.

\begin{figure}
\begin{center}
\includegraphics[width=0.5\textwidth]{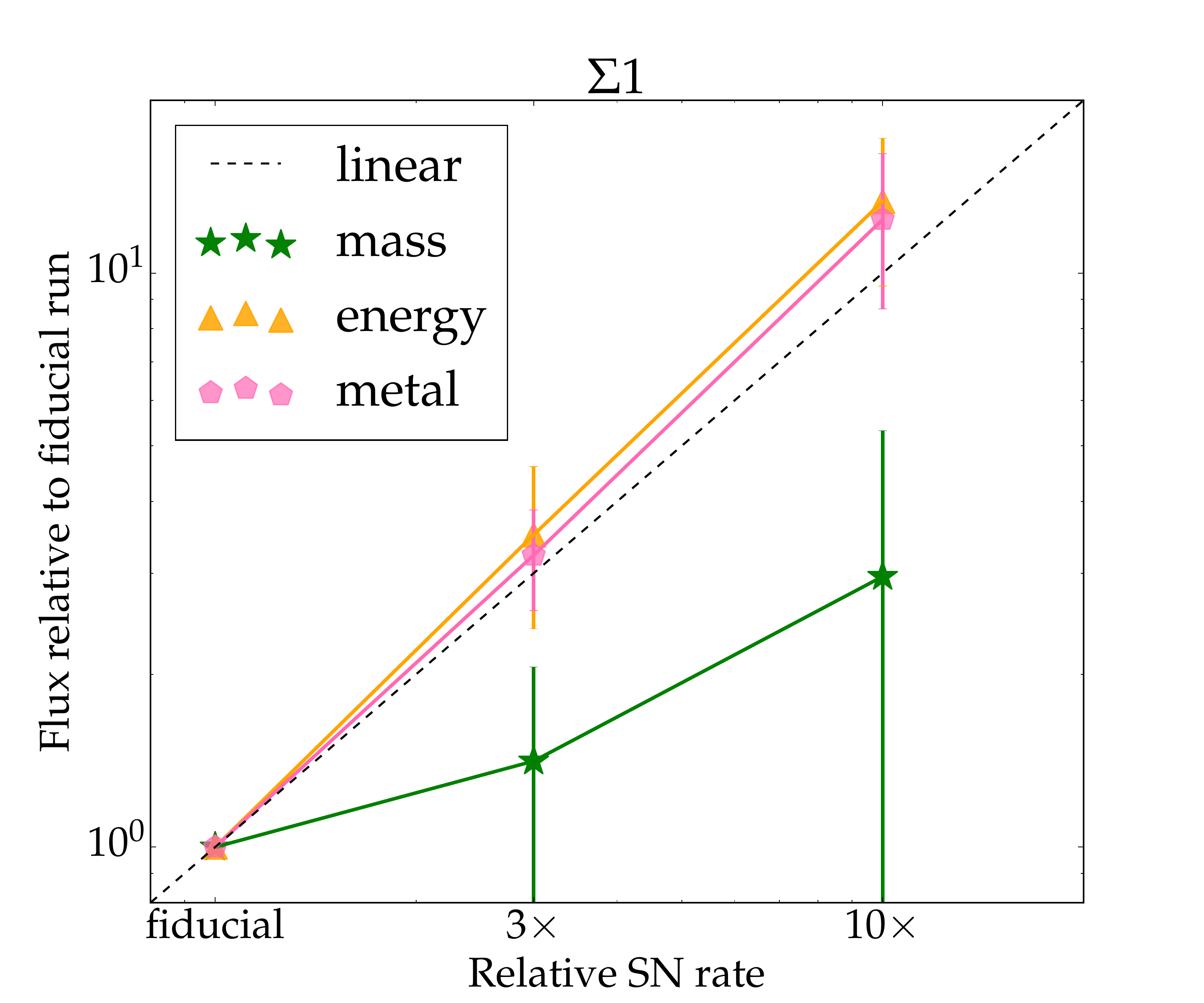}
\caption{Relative fluxes of mass, energy and metal as a function of SN rate for \siggas$ = 1\msun/\pc^2$. The fiducial run follows the KS relation, while the ``enhanced-rate" runs have SN rates increased by 3 and 10 times, respectively. The dashed black line indicates a linear relation.See Section \ref{sec:SN_rate} for discussions.}

\label{f:enhanced_SN_rate}
\end{center}
\end{figure}

We take the run $\Sigma$1-KS, and increase the SN rate by 3$\times$ and 10$\times$, respectively. These runs are listed in Table 1 as ``$\Sigma1$-3KS" and ``$\Sigma1$-10KS". Fig \ref{f:enhanced_SN_rate} shows the mass, energy and metal fluxes of the outflows relative to the fiducial run. The mass flux scales with the SN rate in a sub-linear manner. A factor of 3 and 10 increase in the SN rate only results in, on average, a factor of 1.5 and 3 enhancement in the mass flux, respectively. The energy and metal fluxes, on the other hand, show a roughly linear correlation with the SN rate. This means that the mass loading is less efficient when we increase the SN rate, while the energy and metal loading factors remain roughly constant for different SN rates. 

We caution that, even for the fiducial run, which has the lowest SN rate, most of midplane is in a hot-dominated multiphase state. The sub-linear dependence  of the mass flux, and roughly linear dependence of the energy and metal flux are likely to be the feature in this regime. If, for example, one starts with a SN rate sufficiently small so that the SNRs in the disk would not overlap, then the enhancement of the SN rate would lead to a transition from a steady-state ISM to forming outflows. As a result, the dependence of all fluxes on the SN rate would be super-linear. 
But since we are, in this paper, interested in the regime where outflows are generated, we do not explicitly explore the parameter space that leads to a steady-state ISM. Indeed, for all the four fiducial runs, which are along the KS relation, the ISM at mid-plane is in a thermal runaway state and outflows are being launched. Therefore, the scaling relations as shown in Fig \ref{f:enhanced_SN_rate} should hold for other \siggas cases as well.
 
\cite{girichidis16b} find that clustering of some SNe does not affect the mass outflow rate. We find a very mild increase in mass flux, although with large fluctuations. Within error bars our results are consistent with each other. Overall, once the ISM is hot-dominated, clustering of SNe does not help with the loading factors, and may even be negative for loading mass.

\subsection{Comparison with other works}
\label{sec:compare}

\cite{girichidis16b} have found that for the solar neighbourhood, SNe can blow away most of the gas in the midplane, and drive outflows with a mass loading up to 10. This is higher than our value, which is around 2-3. Compared to our model, their SN scale height is smaller, 50 pc, which causes an ``explosive'' thermal-runaway at the midplane. Initially there is no leak of those hot gas, whose high-pressure propels the neutral gas layer up, like the formation of a super-bubble. Later on, the Rayleigh-Taylor instability will develop, the shell will fragment, and hot gas leaks to form winds \citep{maclow98}. For our case, \hSN\ is larger, meaning that SNe are more spread-out in z-direction. We do see warm shells of gas being driven out initially, but that does not involve too much mass, and would later either go beyond the box, or fragment and fall back. Our loading factors are calculated after those initial transient stage. Additionally, a weaker g-field (see Section \ref{sec:intro_g}) may also partially account for their relatively large mass loading.

\cite{creasey13} study SNe-driven outflows covering a broad parameter space of \siggas and g-field. The g-field can be expressed using the gas fraction,  $f_{\rm{g}} \equiv \Sigma_{\rm{gas}}/(\Sigma_{gas}+\Sigma_*)$. We here conduct on a one-on-one comparison between our fiducial runs and their models. We note that their boxes are smaller in the vertical direction, $|z|\leqslant$ 0.5 kpc, and their outflow fluxes are measured at the outer boundaries; whereas ours are averaged over $1\leqslant|z|\leqslant 2.5$ kpc. Our fiducial runs that overlap with their models are $\Sigma$10-KS, $\Sigma55$-KS, and $\Sigma150$-KS, which corresponds to (\siggas, $f_{\rm{g}}$) = (10, 0.15), (55, 0.5), (150, 0.7), respectively. We convert our g-field from the DM halo to an equivalent surface density of $\approx 25\msun/\pc^2$. We interpolate their data if there is no direct comparison.

For the solar neighbourhood, their mass loading factor is about unity, and energy loading (``thermalization factor'' in their terminology) is around 0.1. These are slightly smaller, by a factor of 1.5-2, compared to our simulation. For the higher \siggas cases, however, the discrepancies are larger. For $\Sigma55$-KS and $\Sigma150$-KS, our \ml are 0.9$\pm$0.3 and 0.2$\pm$0.05, whereas theirs are 0.2 and 0.03, smaller by a factor of 4-7 than our values;  for \el, our values are around 0.2 for both cases, whereas theirs are around 0.05, smaller by a factor of 4.  In a follow-up paper, \cite{creasey15} measure the metal loading efficiency of the outflows for some runs in \citep{creasey13}. Our model parameters only overlap with theirs for the solar neighbourhood, in which we have \mel$=$0.65$\pm$ 0.2 and they have a smaller $\approx 0.2$ . Note also that we both assume a KS relation to relate \siggas and \sigSFR, but we convert the SFR to SN rate by assuming $m_0=150\msun$ (definition of $m_0$ in Section \ref{sec:SN_model}), whereas they have $m_0= 100\msun$. This means the difference is even larger by a factor of 1.5. We attribute the discrepancies mainly to their adoption of a cooling cut-off at $10^4$ K. As a result, the neutral gas layer in their runs is thicker, and most SNe lose their energy there, thus the loading factors are smaller (see Section \ref{sec:PEH} for detailed discussions).

\subsection{A brief summary and some missing physics}
\label{sec:miss_phy}

Under the impact of many SN blast waves, the ISM becomes multiphase. The cooler, denser phase settle down near the midplane, whereas the hotter phases escape and form outflows. There are two regimes of the media: (i) a warm/cool-dominated ISM, where, if a SN explodes within, it would lose most energy by radiative cooling, and (ii) a hot-dominated ISM, where SN shock waves would propagate much faster and further, while the cooling is inefficient \citep{cowie81,li15}. The fraction of SNe that explode in a hot-dominated ISM, \fSN, is key in determining the efficiency of the loading efficiency of energy and metals. A stronger external g-field or a weaker PEH leads to a smaller \hgas, leaving more SNe exploding in low-density medium, thus a more powerful loading of energy and metals; a larger \hSN\ has a similar effect. Simply enhancing the SN rate, without changing \hSN\ or \hgas, yields unchanged energy and metal loading. The mass loading factor, on the other hand, has a more complex dependence on \fSN: a \fSN that is either too small or too large would result in a small mass loading factor. 

We briefly discuss the possible impact of the physics that we do not include in this work. 

\underline{Under-resolved SNe}: 
The resolution and $R_{\rm{inj}}$ for each run are fixed,  and are chosen based on $R_{\rm{cool}}$ for the initial $\rho_{\rm{mid}}$. Later, when the ISM becomes multiphase, SNe exploding in the tenuous/hot phase continue to be well resolved. For all fiducial runs, the hot gas volume fraction at mid-plane is 60-80\%. Since SNe are randomly located, a similar fraction of SNe would explode in the hot phase. The rest SNe which explode in the denser phase are likely under-resolved, but we argue that these SNe are unimportant to drive large-scale outflows. Take $\Sigma$10-KS for an example, the cold phase has a density of about 10 cm$^{-3}$, corresponding to a $R_{\rm{cool}} \sim $ 7 pc. So even when the evolution of the SNR is resolved spatially and temporally, it will lose the majority of its energy at  $\rm R_{\rm{cool}}$. This means that these SNe only have a very localized impact, and contribute little to large-scale outflows.

\underline{Magnetic fields}: In the solar neighbourhood, the magnetic pressure is overall similar to the thermal pressure of gas, and even larger for dense phases \citep{heiles05}. This extra pressure, if included, would be likely to enhance \hgas, and also make SN bubbles smaller \citep{slavin93}. The net effect on the loading factors would be similar to having a stronger PEH -- the energy and metal loadings would be smaller, while the change of the mass loading is not certain. The magnetic field on the dynamics of diffuse ISM is rather mild for the solar neighbourhood \citep{deavillez05, hill12, walch15}, except possibly for providing support for the vertical distribution of the gas \citep{cox05}. Magnetic fields alone are unlikely to play an active role in driving the outflows. 

\underline{Self-gravity}: 
We did not have self-gravity in our simulations. For the solar neighbourhood, only about 1\% of the volume is in a self-gravitating state \citep{draine11}. For larger gas surface density cases, however, the self-gravity is probably more important. Including self-gravity would make the cold phase smaller in volume, thus may facilitate feedback for those SNe that explode in the inter-cloud space \citep[e.g.][]{girichidis16b,kim16}. But we note that (i) the external gravitational field in our simulations does include that from the gas, so the effect of “self-gravity” is not fully missed, at least in the vertical direction. (ii) As mentioned in Section \ref{sec:multiphase}, essentially in all runs, the volume fraction of hot gas at the midplane is 60-80\%, thus we believe shrinking the volume fraction of cold gas by some percentage would not lead to a qualitative change in the loading factors of outflows. (iii) The self-gravity should be included with caution, that is, one should also resolve the counterbalancing force -- feedback acting below the Jeans scale of the dense clumps, which is challenging given the resolution of the current ISM simulations; otherwise, most gas would collapse into a few small clumps, which is also not realistic.

\underline{Cosmic rays (CRs)}: SNe are considered to be the main acceleration sites for CRs. Around 5\%-15\% of SNe energy may go into CRs \citep{hillas05}. Recent simulations show CRs are promising candidates to drive galactic scale outflows, with a mass loading around 0.5 \citep{uhlig12, hanasz13, booth13, salem14}. Recent high-resolution, local simulations indicate that for the solar neighbourhood, both CRs and SN thermal feedback can drive an outflow with mass loading around unity \citep{girichidis16a, simpson16}. CR-driven outflows are cooler, slower, and smoother than the thermally-driven ones. \cite{peters15} point out that compared to the thermal feedback, CR-driven outflows give too little hot gas to match the soft X-ray background. Many questions remain open regarding how CRs propagate in, and interact with, a multiphase ISM. Further work is needed to determine whether thermal feedback or CRs are the dominant force launching outflows. 

\underline{Radiation pressure}: Radiation pressure can propel gas out if gas is optically thick, especially for the infrared light. The photon energy from the stars can then be effectively utilized to drive outflows \citep{murray05,hopkins12}. This can be the case for the extremely dense and dusty SF regions, such as in the ultra-luminous infrared galaxies (ULIRGs). We do not actually explore those extreme cases (Fig \ref{f:bigiel08}). For the \siggas adopted in our models, the optical depth for the infrared is much less than unity, therefore the radiation pressure is not important. 

We also do not include the preprocessing of the ISM by other stellar feedbacks, such as stellar winds, ionizing photons, etc. In general these feedbacks alone do not contribute directly to launching outflows (unless the radiation is highly trapped), but they may make the ISM inhomogeneous. Therefore it is possible for SN to explode in a less dense environment, thus facilitating SN feedback to some extent \citep[e.g.][]{gatto16,peters17}. 

\section{Discussion}
\label{sec:discussion}

\subsection{Hot outflows and hot CGM}
\label{sec:CGM}

Hot, X-ray emitting coronae have been observed around the MW \citep{snowden98} and other star-forming disk galaxies, up to a few tens of kpc away from the galaxy \citep{strickland04,tullmann06, anderson11,dai12}. Hot gas around the MW is also detected through OVII, OVIII absorption lines \citep{fang06, bregman07,gupta12}.The COS-Halos survey finds the warm-hot metal-enriched gas  out to $\gtrsim$ 100 kpc for MW-like galaxies, as traced by the OVI absorption lines \citep{tumlinson11}. Our simulations indicate that the hot phase in the outflows has the largest $v_z$ and $v_{\rm{\widetilde{\mathcal{B}}}}$, and can potentially travel distances comparable to the size of DM halos. Hot outflows carry 10\%-50\% of energy and 30-50\% of metals produced by SNe (Section \ref{sec:load}). Thus, the hot, metal-enriched CGM may at least be partly due to this SNe-driven outflows. Those outflows should have an important dynamic impact on the CGM \citep{hopkins12} .

Recently, \cite{faerman16} create a two-phase phenomenological model of the halo gas for MW-like galaxies, which simultaneously fits the absorption features of OVI, OVII and OVIII. The two phases in their model have volume-weighted medium temperature of $3\times 10^5$ K and $1.8\times 10^6$ K, respectively. Nearly 90\% of the mass is in the hotter phase. The total mass of the two phases are comparable to the baryons in the DM halo but not in the galaxy. Our hot phase in $\Sigma$10-KS and $\Sigma$10-KS-4g has a temperature comparable to that in their model. We note, though, that the cooling luminosity of the CGM in their model is about twice the SNe heating rate from the MW, and given that \el$\approx$0.45 for the MW-average, the total discrepancy is about a factor of 4.

\subsection{Cool outflows}

Warm/cool outflows are observed through interstellar absorption lines, which have velocities of about several hundred km/s \citep{steidel96, shapley03, martin05, weiner09, heckman15}. Mass loading factors around a few are frequently reported. In our simulations, the cool phase in general does not have such high velocities, and the mass loading is usually smaller than unity, especially for high \siggas. We here discuss the possible reasons for this apparent discrepancy. 

We first note that observations usually focus on the star-bursting regime, such as (U)LIRGs and Lyman break galaxies, which we do not cover in our simulations (Fig \ref{f:bigiel08}). As mentioned before, radiation pressure may account for the potential high mass loading in those extreme environments, provided that the infrared light can be trapped by dust. Note that those bursting SF systems are very rare;  even at higher redshifts when they are more common, they only account for $\lesssim$ 10\% of cosmic SF density \citep{rodighiero11}. 

Second, the observed mass loading factors have large uncertainties, due to the unconstrained metallicity, ionization fraction, geometry, etc. This can lead to an error bar as large as the mass loading factor itself. Recent work by \cite{chisholm16}, which claims a better constraints on the above quantities, suggests a small \ml$\approx$ 0.1. This is, in fact, not inconsistent with our value for the highest \siggas. 

On the simulation side, we note that even though we adopt high enough resolution to make sure the Sedov-Taylor phase of SNRs is well-resolved,  the interaction between different gas phases may still not be sufficiently resolved, once the ISM becomes multiphase. To what extent this may affect the mass loading is not clear.  

Another apparent discrepancy is that we find that the cooler phases dominate less of the outflow mass flux for higher \sigSFR. It is possible, however, that some cool clouds may form on the way due to thermal instability \citep{field65,thompson16}, which we do not capture because of the relatively small box size. But again, how much cool mass would form under this mechanism is not clear.

Overall, there remains much uncertainty about the observational interpretation and theoretical modelling of cool outflows.

\subsection{Implications for cosmological simulations}

Stellar feedback is one of the key ingredients for galaxy formation in cosmological simulations and semi-analytic models \citep{somerville15}. The general consensus is that strong feedback/outflows are needed to reproduce various observed scaling relations of galaxies \citep{springel03, stinson06, hopkins12, hummels12}. Due to the limited resolutions in cosmological simulations, however, multiphase ISM and individual SNR cannot be resolved. Ad hoc models are widely used in the field, with free parameters fine-tuned to match observations. Different groups adopt very different recipes, usually invoking shutting off cooling or hydrodynamics. To evaluate the real impact of feedback, however, a physically-grounded model is necessary. While formulating such a model is beyond the scope of this paper, we compare the loading factors found in our work to current cosmological simulations, and briefly discuss the implications.

In cosmological simulations, a significant fraction of SNe energy is used to generate outflows, roughly 30-50\% \citep[e.g.][]{oppenheimer08}. Our \el varies from 15-50\%, broadly consistent with that fraction. To match the observed mass-matallicity relation of galaxies, a significant fraction of metals have to be driven out of the galaxies. Our \mel is about 40-80\%, in general agreement with what observations require and cosmological simulation adopt. The main difference is the mass loading of the outflows, or in other words, the amount of mass that those energy and metals couple to. Our \ml is in general smaller than what is adopted in cosmological simulations, by roughly an order of magnitude. A smaller mass loading is not surprising for the simulations where the multiphase ISM is resolved: SN blast waves tend to vent through the least-obstructed channel, thus preferentially heating and accelerating the tenuous phase \citep{cowie81, li15}. Cosmological simulations cannot resolve the multiphase ISM/outflows generally. Mixing the fast/tenuous and the slow/dense phase due to insufficient resolution would result in a slower, cooler and mass-loaded outflows. Therefore current cosmological simulations may not accurately predict the impacts of galactic outflows on the CGM and galaxy formation. From our simulations, hot outflows, though not carrying a great amount of mass, may be able to suppress the inflows given their vigor (Section \ref{sec:vel}), therefore restricting galaxy masses. The implications of tenuous yet vigorous hot outflows to galaxy formation and the CGM are not clear (although see recent attempts by \cite{dave16}, \cite{fielding17}).

\section{Summary}

In this paper, we use high-resolution simulations to study the multiphase outflows driven by supernovae from stratified media. We cover a wide range of \siggas: 1-150$\msun/\rm{pc}^2$. We quantify the multiphase outflows by measuring the loading factors of energy, mass and metals. The fiducial runs assume \siggas scales with \sigSFR\ as in the Kennicutt-Schmidt relation (Fig \ref{f:bigiel08}). We study the effects of various physics on the loading factors: SN scale height \hSN, photoelectric heating, external gravitational field, and enhanced SN rate.
Here are our main conclusions:

1. The ISM quickly becomes multiphase under the impact of multiple SNe. The cold phase settles down near the midplane, whereas hotter phases preferentially escape and form outflows.

2. For the solar neighbourhood case, the gas pressure, volume fraction of hot gas, and mean densities of different gas phases agree well with the observations.

3. The mass loading factor \ml decreases monotonically with increasing \siggas as \ml$\propto \Sigma^{-0.6}_{\rm{gas}}$ (Eq. \ref{eq:ml}). The outflowing flux is about 0.1-10 of \sigSFR. The energy and metal loading factors do not show significant variance with \siggas. Roughly 10-50\% of the energy and 40-80\% of the metals produced by SNe are carried away by the outflows (Fig \ref{f:loading_all}).

4. More of the outflow volume is occupied by hot gas ($T > 3\times 10^5$ K) for larger \siggas. The hot phase contributes to $\gtrsim$1/3 of the mass loading, $>$0.8 of the energy loading, and 0.5-0.9 of the metal loading (Fig \ref{f:loading_all} ). It has significantly larger $v_z$ and \vB (see Section \ref{sec:vel} for definition) than the cooler phases. Hot outflows are very likely to have a broad impact on the CGM.

5. Increasing $h_{\rm{SN}}$ enhances the energy and metal loading, since more energy/metals are directly deposited into the low-density halo. The mass loading factor, on the other hand, does not show a monotonic dependence on $h_{\rm{SN}}$. The relative scale height of SNe and gas is a very important factor determining the loading efficiencies. Various physical processes affect the loading factors by changing $h_{\rm{SN}}$ and/or $h_{\rm{gas}}$.

6. A stronger PEH makes SN feedback less effective, since it keeps more gas in the warm phase, thus a larger scale height of the neutral gas layer. In the extreme case where the cooling curve has a lower cut-off at $10^4$ K, the feedback is severely suppressed, with the loading factors smaller by a factor of $\gtrsim$ 10. (Fig \ref{f:300_1e4K}, \ref{f:loading_300_1e4K}). 

7. A larger gravitational field, by lowering \hSN, may result in much stronger energy and metal loading (Fig \ref{f:g_field}). 

8. If the SN rate is enhanced above the Kennicutt relation, the energy and metal fluxes roughly scale linearly with the SN rate, but the mass flux has a sub-linear dependence on the SN rate. Overall, once the ISM is hot-dominated, clustering SNe does not enhance the time-integrated properties of outflows.

\section*{Acknowledgement}
We thank the referee for comments that help to improve the clarity of the paper. We thank Thorsten Naab for very helpful comments on the manuscript, and Eve Ostriker for interesting discussions on the topic. Computations described in this work were performed using the publicly-available \texttt{Enzo} code (http://enzo-project.org), which is the product of a collaborative effort of many independent scientists from numerous institutions around the world.  Their commitment to open science has helped make this work possible. The visualization is partly done using the \texttt{yt} project \citep{turk11}. This work utilized the following computing resources: the Extreme Science and Engineering Discovery Environment (XSEDE) supported by National Science Foundation grant number ACI-1053575, the NASA HECC Pleiades Supercomputer (under accounts of s1670 and s1529), the Hecate cluster at Princeton University, and the Yeti cluster at Columbia University. The authors acknowledge financial support from NASA grant NNX15AB20G and NSF grants AST-1312888 and AST-1615955.

%\vspace{1in}

%\bibliography{master_bib}

\begin{thebibliography}{}
\expandafter\ifx\csname natexlab\endcsname\relax\def\natexlab#1{#1}\fi

\bibitem[{{Anderson} \& {Bregman}(2011)}]{anderson11}
{Anderson}, M.~E., \& {Bregman}, J.~N. 2011, \apj, 737, 22

\bibitem[{{Bigiel} {et~al.}(2008){Bigiel}, {Leroy}, {Walter}, {Brinks}, {de
  Blok}, {Madore}, \& {Thornley}}]{bigiel08}
{Bigiel}, F., {Leroy}, A., {Walter}, F., {et~al.} 2008, \aj, 136, 2846

\bibitem[{{Binney} \& {Tremaine}(2008)}]{binney08}
{Binney}, J., \& {Tremaine}, S. 2008, {Galactic Dynamics: Second Edition}
  (Princeton University Press)

\bibitem[{{Bolatto} {et~al.}(2013){Bolatto}, {Warren}, {Leroy}, {Walter},
  {Veilleux}, {Ostriker}, {Ott}, {Zwaan}, {Fisher}, {Weiss}, {Rosolowsky}, \&
  {Hodge}}]{bolatto13}
{Bolatto}, A.~D., {Warren}, S.~R., {Leroy}, A.~K., {et~al.} 2013, \nat, 499,
  450

\bibitem[{{Booth} {et~al.}(2013){Booth}, {Agertz}, {Kravtsov}, \&
  {Gnedin}}]{booth13}
{Booth}, C.~M., {Agertz}, O., {Kravtsov}, A.~V., \& {Gnedin}, N.~Y. 2013,
  \apjl, 777, L16

\bibitem[{{Bregman} \& {Lloyd-Davies}(2007)}]{bregman07}
{Bregman}, J.~N., \& {Lloyd-Davies}, E.~J. 2007, \apj, 669, 990

\bibitem[{{Bryan} {et~al.}(2014){Bryan}, {Norman}, {O'Shea}, {Abel}, {Wise},
  {Turk}, {Reynolds}, {Collins}, {Wang}, {Skillman}, {Smith}, {Harkness},
  {Bordner}, {Kim}, {Kuhlen}, {Xu}, {Goldbaum}, {Hummels}, {Kritsuk}, {Tasker},
  {Skory}, {Simpson}, {Hahn}, {Oishi}, {So}, {Zhao}, {Cen}, {Li}, \& {Enzo
  Collaboration}}]{bryan14}
{Bryan}, G.~L., {Norman}, M.~L., {O'Shea}, B.~W., {et~al.} 2014, \apjs, 211, 19

\bibitem[{{Bustard} {et~al.}(2016){Bustard}, {Zweibel}, \&
  {D'Onghia}}]{bustard16}
{Bustard}, C., {Zweibel}, E.~G., \& {D'Onghia}, E. 2016, \apj, 819, 29

\bibitem[{{Chen} {et~al.}(2010){Chen}, {Tremonti}, {Heckman}, {Kauffmann},
  {Weiner}, {Brinchmann}, \& {Wang}}]{chen10}
{Chen}, Y.-M., {Tremonti}, C.~A., {Heckman}, T.~M., {et~al.} 2010, \aj, 140,
  445

\bibitem[{{Chisholm} {et~al.}(2016){Chisholm}, {Tremonti}, {Leitherer}, \&
  {Chen}}]{chisholm16}
{Chisholm}, J., {Tremonti}, C.~A., {Leitherer}, C., \& {Chen}, Y. 2016, ArXiv
  e-prints, arXiv:1605.05769

\bibitem[{{Colella} \& {Woodward}(1984)}]{colella84}
{Colella}, P., \& {Woodward}, P.~R. 1984, Journal of Computational Physics, 54,
  174

\bibitem[{{Cowie} {et~al.}(1981){Cowie}, {McKee}, \& {Ostriker}}]{cowie81}
{Cowie}, L.~L., {McKee}, C.~F., \& {Ostriker}, J.~P. 1981, \apj, 247, 908

\bibitem[{{Cox}(2005)}]{cox05}
{Cox}, D.~P. 2005, \araa, 43, 337

\bibitem[{{Cox} \& {Smith}(1974)}]{cox74}
{Cox}, D.~P., \& {Smith}, B.~W. 1974, \apjl, 189, L105

\bibitem[{{Creasey} {et~al.}(2015){Creasey}, {Scannapieco}, {Nuza}, {Yepes},
  {Gottl{\"o}ber}, \& {Steinmetz}}]{creasey15}
{Creasey}, P., {Scannapieco}, C., {Nuza}, S.~E., {et~al.} 2015, \apjl, 800, L4

\bibitem[{{Creasey} {et~al.}(2013){Creasey}, {Theuns}, \& {Bower}}]{creasey13}
{Creasey}, P., {Theuns}, T., \& {Bower}, R.~G. 2013, \mnras, 429, 1922

\bibitem[{{Dai} {et~al.}(2012){Dai}, {Anderson}, {Bregman}, \&
  {Miller}}]{dai12}
{Dai}, X., {Anderson}, M.~E., {Bregman}, J.~N., \& {Miller}, J.~M. 2012, \apj,
  755, 107

\bibitem[{{Dav{\'e}} {et~al.}(2016){Dav{\'e}}, {Thompson}, \&
  {Hopkins}}]{dave16}
{Dav{\'e}}, R., {Thompson}, R., \& {Hopkins}, P.~F. 2016, \mnras, 462, 3265

\bibitem[{{de Avillez}(2000)}]{deavillez00}
{de Avillez}, M.~A. 2000, \mnras, 315, 479

\bibitem[{{de Avillez} \& {Breitschwerdt}(2005)}]{deavillez05}
{de Avillez}, M.~A., \& {Breitschwerdt}, D. 2005, \aap, 436, 585

\bibitem[{{Draine}(1978)}]{draine78}
{Draine}, B.~T. 1978, \apjs, 36, 595

\bibitem[{{Draine}(2011)}]{draine11}
---. 2011, {Physics of the Interstellar and Intergalactic Medium}

\bibitem[{{Efstathiou}(2000)}]{efstathiou00}
{Efstathiou}, G. 2000, \mnras, 317, 697

\bibitem[{{Erb} {et~al.}(2006){Erb}, {Shapley}, {Pettini}, {Steidel}, {Reddy},
  \& {Adelberger}}]{erb06}
{Erb}, D.~K., {Shapley}, A.~E., {Pettini}, M., {et~al.} 2006, \apj, 644, 813

\bibitem[{{Faerman} {et~al.}(2016){Faerman}, {Sternberg}, \&
  {McKee}}]{faerman16}
{Faerman}, Y., {Sternberg}, A., \& {McKee}, C.~F. 2016, ArXiv e-prints,
  arXiv:1602.00689

\bibitem[{{Fang} {et~al.}(2006){Fang}, {Mckee}, {Canizares}, \&
  {Wolfire}}]{fang06}
{Fang}, T., {Mckee}, C.~F., {Canizares}, C.~R., \& {Wolfire}, M. 2006, \apj,
  644, 174

\bibitem[{{Field}(1965)}]{field65}
{Field}, G.~B. 1965, \apj, 142, 531

\bibitem[{{Fielding} {et~al.}(2017){Fielding}, {Quataert}, {McCourt}, \&
  {Thompson}}]{fielding17}
{Fielding}, D., {Quataert}, E., {McCourt}, M., \& {Thompson}, T.~A. 2017,
  \mnras, 466, 3810

\bibitem[{{Freeman}(1987)}]{freeman87}
{Freeman}, K.~C. 1987, \araa, 25, 603

\bibitem[{{Fujita} {et~al.}(2004){Fujita}, {Mac Low}, {Ferrara}, \&
  {Meiksin}}]{fujita04}
{Fujita}, A., {Mac Low}, M.-M., {Ferrara}, A., \& {Meiksin}, A. 2004, \apj,
  613, 159

\bibitem[{{Gatto} {et~al.}(2015){Gatto}, {Walch}, {Low}, {Naab}, {Girichidis},
  {Glover}, {W{\"u}nsch}, {Klessen}, {Clark}, {Baczynski}, {Peters},
  {Ostriker}, {Ib{\'a}{\~n}ez-Mej{\'{\i}}a}, \& {Haid}}]{gatto15}
{Gatto}, A., {Walch}, S., {Low}, M.-M.~M., {et~al.} 2015, \mnras, 449, 1057

\bibitem[{{Gatto} {et~al.}(2016){Gatto}, {Walch}, {Naab}, {Girichidis},
  {W{\"u}nsch}, {Glover}, {Klessen}, {Clark}, {Peters}, {Derigs}, {Baczynski},
  \& {Puls}}]{gatto16}
{Gatto}, A., {Walch}, S., {Naab}, T., {et~al.} 2016, ArXiv e-prints,
  arXiv:1606.05346

\bibitem[{{Genel} {et~al.}(2015){Genel}, {Fall}, {Hernquist}, {Vogelsberger},
  {Snyder}, {Rodriguez-Gomez}, {Sijacki}, \& {Springel}}]{genel15}
{Genel}, S., {Fall}, S.~M., {Hernquist}, L., {et~al.} 2015, \apjl, 804, L40

\bibitem[{{Gent} {et~al.}(2013){Gent}, {Shukurov}, {Fletcher}, {Sarson}, \&
  {Mantere}}]{gent13}
{Gent}, F.~A., {Shukurov}, A., {Fletcher}, A., {Sarson}, G.~R., \& {Mantere},
  M.~J. 2013, \mnras, 432, 1396

\bibitem[{{Genzel} {et~al.}(2011){Genzel}, {Newman}, {Jones}, {F{\"o}rster
  Schreiber}, {Shapiro}, {Genel}, {Lilly}, {Renzini}, {Tacconi}, {Bouch{\'e}},
  {Burkert}, {Cresci}, {Buschkamp}, {Carollo}, {Ceverino}, {Davies}, {Dekel},
  {Eisenhauer}, {Hicks}, {Kurk}, {Lutz}, {Mancini}, {Naab}, {Peng},
  {Sternberg}, {Vergani}, \& {Zamorani}}]{genzel11}
{Genzel}, R., {Newman}, S., {Jones}, T., {et~al.} 2011, \apj, 733, 101

\bibitem[{{Gilmore} \& {Reid}(1983)}]{gilmore83}
{Gilmore}, G., \& {Reid}, N. 1983, \mnras, 202, 1025

\bibitem[{{Girichidis} {et~al.}(2016{\natexlab{a}}){Girichidis}, {Naab},
  {Walch}, {Hanasz}, {Mac Low}, {Ostriker}, {Gatto}, {Peters}, {W{\"u}nsch},
  {Glover}, {Klessen}, {Clark}, \& {Baczynski}}]{girichidis16a}
{Girichidis}, P., {Naab}, T., {Walch}, S., {et~al.} 2016{\natexlab{a}}, \apjl,
  816, L19

\bibitem[{{Girichidis} {et~al.}(2016{\natexlab{b}}){Girichidis}, {Walch},
  {Naab}, {Gatto}, {W{\"u}nsch}, {Glover}, {Klessen}, {Clark}, {Peters},
  {Derigs}, \& {Baczynski}}]{girichidis16b}
{Girichidis}, P., {Walch}, S., {Naab}, T., {et~al.} 2016{\natexlab{b}}, \mnras,
  456, 3432

\bibitem[{{Governato} {et~al.}(2007){Governato}, {Willman}, {Mayer}, {Brooks},
  {Stinson}, {Valenzuela}, {Wadsley}, \& {Quinn}}]{governato07}
{Governato}, F., {Willman}, B., {Mayer}, L., {et~al.} 2007, \mnras, 374, 1479

\bibitem[{{Gupta} {et~al.}(2012){Gupta}, {Mathur}, {Krongold}, {Nicastro}, \&
  {Galeazzi}}]{gupta12}
{Gupta}, A., {Mathur}, S., {Krongold}, Y., {Nicastro}, F., \& {Galeazzi}, M.
  2012, \apjl, 756, L8

\bibitem[{{Hanasz} {et~al.}(2013){Hanasz}, {Lesch}, {Naab}, {Gawryszczak},
  {Kowalik}, \& {W{\'o}lta{\'n}ski}}]{hanasz13}
{Hanasz}, M., {Lesch}, H., {Naab}, T., {et~al.} 2013, \apjl, 777, L38

\bibitem[{{Heckman} {et~al.}(2015){Heckman}, {Alexandroff}, {Borthakur},
  {Overzier}, \& {Leitherer}}]{heckman15}
{Heckman}, T.~M., {Alexandroff}, R.~M., {Borthakur}, S., {Overzier}, R., \&
  {Leitherer}, C. 2015, \apj, 809, 147

\bibitem[{{Heckman} {et~al.}(2001){Heckman}, {Sembach}, {Meurer}, {Strickland},
  {Martin}, {Calzetti}, \& {Leitherer}}]{heckman01}
{Heckman}, T.~M., {Sembach}, K.~R., {Meurer}, G.~R., {et~al.} 2001, \apj, 554,
  1021

\bibitem[{{Heiderman} {et~al.}(2010){Heiderman}, {Evans}, {Allen}, {Huard}, \&
  {Heyer}}]{heiderman10}
{Heiderman}, A., {Evans}, II, N.~J., {Allen}, L.~E., {Huard}, T., \& {Heyer},
  M. 2010, \apj, 723, 1019

\bibitem[{{Heiles} \& {Crutcher}(2005)}]{heiles05}
{Heiles}, C., \& {Crutcher}, R. 2005, in Lecture Notes in Physics, Berlin
  Springer Verlag, Vol. 664, Cosmic Magnetic Fields, ed. R.~{Wielebinski} \&
  R.~{Beck}, 137

\bibitem[{{Heiles} \& {Troland}(2003)}]{heiles03}
{Heiles}, C., \& {Troland}, T.~H. 2003, \apj, 586, 1067

\bibitem[{{Hennebelle} \& {Iffrig}(2014)}]{hennebelle14}
{Hennebelle}, P., \& {Iffrig}, O. 2014, \aap, 570, A81

\bibitem[{{Hill} {et~al.}(2012){Hill}, {Joung}, {Mac Low}, {Benjamin},
  {Haffner}, {Klingenberg}, \& {Waagan}}]{hill12}
{Hill}, A.~S., {Joung}, M.~R., {Mac Low}, M.-M., {et~al.} 2012, \apj, 750, 104

\bibitem[{{Hillas}(2005)}]{hillas05}
{Hillas}, A.~M. 2005, Journal of Physics G Nuclear Physics, 31, R95

\bibitem[{{Hopkins} {et~al.}(2012){Hopkins}, {Quataert}, \&
  {Murray}}]{hopkins12}
{Hopkins}, P.~F., {Quataert}, E., \& {Murray}, N. 2012, \mnras, 421, 3522

\bibitem[{{Hummels} \& {Bryan}(2012)}]{hummels12}
{Hummels}, C.~B., \& {Bryan}, G.~L. 2012, \apj, 749, 140

\bibitem[{{Joung} \& {Mac Low}(2006)}]{joung06}
{Joung}, M.~K.~R., \& {Mac Low}, M.-M. 2006, \apj, 653, 1266

\bibitem[{{Joung} {et~al.}(2009){Joung}, {Mac Low}, \& {Bryan}}]{joung09}
{Joung}, M.~R., {Mac Low}, M.-M., \& {Bryan}, G.~L. 2009, \apj, 704, 137

\bibitem[{{Kim} \& {Ostriker}(2015)}]{kim15}
{Kim}, C.-G., \& {Ostriker}, E.~C. 2015, \apj, 802, 99

\bibitem[{{Kim} \& {Ostriker}(2016)}]{kim16}
---. 2016, ArXiv e-prints, arXiv:1612.03918

\bibitem[{{Kuijken} \& {Gilmore}(1989)}]{kuijken89}
{Kuijken}, K., \& {Gilmore}, G. 1989, \mnras, 239, 605

\bibitem[{{Li} {et~al.}(2015){Li}, {Ostriker}, {Cen}, {Bryan}, \&
  {Naab}}]{li15}
{Li}, M., {Ostriker}, J.~P., {Cen}, R., {Bryan}, G.~L., \& {Naab}, T. 2015,
  \apj, 814, 4

\bibitem[{{Mac Low} \& {Ferrara}(1998)}]{maclow98}
{Mac Low}, M.-M., \& {Ferrara}, A. 1998, in Lecture Notes in Physics, Berlin
  Springer Verlag, Vol. 506, IAU Colloq. 166: The Local Bubble and Beyond, ed.
  D.~{Breitschwerdt}, M.~J. {Freyberg}, \& J.~{Truemper}, 559--562

\bibitem[{{Mac Low} \& {Ferrara}(1999)}]{maclow99}
{Mac Low}, M.-M., \& {Ferrara}, A. 1999, \apj, 513, 142

\bibitem[{{Martin}(2005)}]{martin05}
{Martin}, C.~L. 2005, \apj, 621, 227

\bibitem[{{McKee} \& {Ostriker}(1977)}]{mckee77}
{McKee}, C.~F., \& {Ostriker}, J.~P. 1977, \apj, 218, 148

\bibitem[{{Mitchell} {et~al.}(1976){Mitchell}, {Culhane}, {Davison}, \&
  {Ives}}]{mitchell76}
{Mitchell}, R.~J., {Culhane}, J.~L., {Davison}, P.~J.~N., \& {Ives}, J.~C.
  1976, \mnras, 175, 29P

\bibitem[{{Murray} {et~al.}(2005){Murray}, {Quataert}, \&
  {Thompson}}]{murray05}
{Murray}, N., {Quataert}, E., \& {Thompson}, T.~A. 2005, \apj, 618, 569

\bibitem[{{Narayan} \& {Ostriker}(1990)}]{narayan90}
{Narayan}, R., \& {Ostriker}, J.~P. 1990, \apj, 352, 222

\bibitem[{{Navarro} {et~al.}(1997){Navarro}, {Frenk}, \& {White}}]{navarro97}
{Navarro}, J.~F., {Frenk}, C.~S., \& {White}, S.~D.~M. 1997, \apj, 490, 493

\bibitem[{{Oppenheimer} \& {Dav{\'e}}(2006)}]{oppenheimer06}
{Oppenheimer}, B.~D., \& {Dav{\'e}}, R. 2006, \mnras, 373, 1265

\bibitem[{{Oppenheimer} \& {Dav{\'e}}(2008)}]{oppenheimer08}
---. 2008, \mnras, 387, 577

\bibitem[{{Peters} {et~al.}(2015){Peters}, {Girichidis}, {Gatto}, {Naab},
  {Walch}, {W{\"u}nsch}, {Glover}, {Clark}, {Klessen}, \&
  {Baczynski}}]{peters15}
{Peters}, T., {Girichidis}, P., {Gatto}, A., {et~al.} 2015, \apjl, 813, L27

\bibitem[{{Peters} {et~al.}(2017){Peters}, {Naab}, {Walch}, {Glover},
  {Girichidis}, {Pellegrini}, {Klessen}, {W{\"u}nsch}, {Gatto}, \&
  {Baczynski}}]{peters17}
{Peters}, T., {Naab}, T., {Walch}, S., {et~al.} 2017, \mnras, 466, 3293

\bibitem[{{Rodighiero} {et~al.}(2011){Rodighiero}, {Daddi}, {Baronchelli},
  {Cimatti}, {Renzini}, {Aussel}, {Popesso}, {Lutz}, {Andreani}, {Berta},
  {Cava}, {Elbaz}, {Feltre}, {Fontana}, {F{\"o}rster Schreiber},
  {Franceschini}, {Genzel}, {Grazian}, {Gruppioni}, {Ilbert}, {Le Floch},
  {Magdis}, {Magliocchetti}, {Magnelli}, {Maiolino}, {McCracken}, {Nordon},
  {Poglitsch}, {Santini}, {Pozzi}, {Riguccini}, {Tacconi}, {Wuyts}, \&
  {Zamorani}}]{rodighiero11}
{Rodighiero}, G., {Daddi}, E., {Baronchelli}, I., {et~al.} 2011, \apjl, 739,
  L40

\bibitem[{{Rosen} \& {Bregman}(1995)}]{rosen95}
{Rosen}, A., \& {Bregman}, J.~N. 1995, \apj, 440, 634

\bibitem[{{Rupke} {et~al.}(2002){Rupke}, {Veilleux}, \& {Sanders}}]{rupke02}
{Rupke}, D.~S., {Veilleux}, S., \& {Sanders}, D.~B. 2002, \apj, 570, 588

\bibitem[{{Salem} \& {Bryan}(2014)}]{salem14}
{Salem}, M., \& {Bryan}, G.~L. 2014, \mnras, 437, 3312

\bibitem[{{Scannapieco} {et~al.}(2008){Scannapieco}, {Tissera}, {White}, \&
  {Springel}}]{scannapieco08}
{Scannapieco}, C., {Tissera}, P.~B., {White}, S.~D.~M., \& {Springel}, V. 2008,
  \mnras, 389, 1137

\bibitem[{{Schaye} {et~al.}(2003){Schaye}, {Aguirre}, {Kim}, {Theuns}, {Rauch},
  \& {Sargent}}]{schaye03}
{Schaye}, J., {Aguirre}, A., {Kim}, T.-S., {et~al.} 2003, \apj, 596, 768

\bibitem[{{Shapley} {et~al.}(2003){Shapley}, {Steidel}, {Pettini}, \&
  {Adelberger}}]{shapley03}
{Shapley}, A.~E., {Steidel}, C.~C., {Pettini}, M., \& {Adelberger}, K.~L. 2003,
  \apj, 588, 65

\bibitem[{{Simpson} {et~al.}(2014){Simpson}, {Bryan}, {Hummels}, \&
  {Ostriker}}]{simpson14}
{Simpson}, C.~M., {Bryan}, G.~L., {Hummels}, C., \& {Ostriker}, J.~P. 2014,
  ArXiv e-prints, arXiv:1410.3822

\bibitem[{{Simpson} {et~al.}(2016){Simpson}, {Pakmor}, {Marinacci}, {Pfrommer},
  {Springel}, {Glover}, {Clark}, \& {Smith}}]{simpson16}
{Simpson}, C.~M., {Pakmor}, R., {Marinacci}, F., {et~al.} 2016, ArXiv e-prints,
  arXiv:1606.02324

\bibitem[{{Slavin} \& {Cox}(1993)}]{slavin93}
{Slavin}, J.~D., \& {Cox}, D.~P. 1993, \apj, 417, 187

\bibitem[{{Snowden} {et~al.}(1998){Snowden}, {Egger}, {Finkbeiner}, {Freyberg},
  \& {Plucinsky}}]{snowden98}
{Snowden}, S.~L., {Egger}, R., {Finkbeiner}, D.~P., {Freyberg}, M.~J., \&
  {Plucinsky}, P.~P. 1998, \apj, 493, 715

\bibitem[{{Somerville} \& {Dav{\'e}}(2015)}]{somerville15}
{Somerville}, R.~S., \& {Dav{\'e}}, R. 2015, \araa, 53, 51

\bibitem[{{Songaila} \& {Cowie}(1996)}]{songaila96}
{Songaila}, A., \& {Cowie}, L.~L. 1996, \aj, 112, 335

\bibitem[{{Springel} \& {Hernquist}(2003)}]{springel03}
{Springel}, V., \& {Hernquist}, L. 2003, \mnras, 339, 312

\bibitem[{{Steidel} {et~al.}(1996){Steidel}, {Giavalisco}, {Pettini},
  {Dickinson}, \& {Adelberger}}]{steidel96}
{Steidel}, C.~C., {Giavalisco}, M., {Pettini}, M., {Dickinson}, M., \&
  {Adelberger}, K.~L. 1996, \apjl, 462, L17

\bibitem[{{Stinson} {et~al.}(2006){Stinson}, {Seth}, {Katz}, {Wadsley},
  {Governato}, \& {Quinn}}]{stinson06}
{Stinson}, G., {Seth}, A., {Katz}, N., {et~al.} 2006, \mnras, 373, 1074

\bibitem[{{Strickland} {et~al.}(2004){Strickland}, {Heckman}, {Colbert},
  {Hoopes}, \& {Weaver}}]{strickland04}
{Strickland}, D.~K., {Heckman}, T.~M., {Colbert}, E.~J.~M., {Hoopes}, C.~G., \&
  {Weaver}, K.~A. 2004, \apjs, 151, 193

\bibitem[{{Thompson} {et~al.}(2016){Thompson}, {Quataert}, {Zhang}, \&
  {Weinberg}}]{thompson16}
{Thompson}, T.~A., {Quataert}, E., {Zhang}, D., \& {Weinberg}, D.~H. 2016,
  \mnras, 455, 1830

\bibitem[{{Tremonti} {et~al.}(2004){Tremonti}, {Heckman}, {Kauffmann},
  {Brinchmann}, {Charlot}, {White}, {Seibert}, {Peng}, {Schlegel}, {Uomoto},
  {Fukugita}, \& {Brinkmann}}]{tremonti04}
{Tremonti}, C.~A., {Heckman}, T.~M., {Kauffmann}, G., {et~al.} 2004, \apj, 613,
  898

\bibitem[{{T{\"u}llmann} {et~al.}(2006){T{\"u}llmann}, {Pietsch}, {Rossa},
  {Breitschwerdt}, \& {Dettmar}}]{tullmann06}
{T{\"u}llmann}, R., {Pietsch}, W., {Rossa}, J., {Breitschwerdt}, D., \&
  {Dettmar}, R.-J. 2006, \aap, 448, 43

\bibitem[{{Tumlinson} {et~al.}(2011){Tumlinson}, {Thom}, {Werk}, {Prochaska},
  {Tripp}, {Weinberg}, {Peeples}, {O'Meara}, {Oppenheimer}, {Meiring}, {Katz},
  {Dav{\'e}}, {Ford}, \& {Sembach}}]{tumlinson11}
{Tumlinson}, J., {Thom}, C., {Werk}, J.~K., {et~al.} 2011, Science, 334, 948

\bibitem[{{Turk} {et~al.}(2011){Turk}, {Smith}, {Oishi}, {Skory}, {Skillman},
  {Abel}, \& {Norman}}]{turk11}
{Turk}, M.~J., {Smith}, B.~D., {Oishi}, J.~S., {et~al.} 2011, \apjs, 192, 9

\bibitem[{{Turner} {et~al.}(2015){Turner}, {Schaye}, {Steidel}, {Rudie}, \&
  {Strom}}]{turner15}
{Turner}, M.~L., {Schaye}, J., {Steidel}, C.~C., {Rudie}, G.~C., \& {Strom},
  A.~L. 2015, \mnras, 450, 2067

\bibitem[{{Uhlig} {et~al.}(2012){Uhlig}, {Pfrommer}, {Sharma}, {Nath},
  {En{\ss}lin}, \& {Springel}}]{uhlig12}
{Uhlig}, M., {Pfrommer}, C., {Sharma}, M., {et~al.} 2012, \mnras, 423, 2374

\bibitem[{{Veilleux} {et~al.}(2005){Veilleux}, {Cecil}, \&
  {Bland-Hawthorn}}]{veilleux05}
{Veilleux}, S., {Cecil}, G., \& {Bland-Hawthorn}, J. 2005, \araa, 43, 769

\bibitem[{{Walch} {et~al.}(2015){Walch}, {Girichidis}, {Naab}, {Gatto},
  {Glover}, {W{\"u}nsch}, {Klessen}, {Clark}, {Peters}, {Derigs}, \&
  {Baczynski}}]{walch15}
{Walch}, S., {Girichidis}, P., {Naab}, T., {et~al.} 2015, \mnras, 454, 238

\bibitem[{{Walter} {et~al.}(2002){Walter}, {Weiss}, \& {Scoville}}]{walter02}
{Walter}, F., {Weiss}, A., \& {Scoville}, N. 2002, \apjl, 580, L21

\bibitem[{{Weiner} {et~al.}(2009){Weiner}, {Coil}, {Prochaska}, {Newman},
  {Cooper}, {Bundy}, {Conselice}, {Dutton}, {Faber}, {Koo}, {Lotz}, {Rieke}, \&
  {Rubin}}]{weiner09}
{Weiner}, B.~J., {Coil}, A.~L., {Prochaska}, J.~X., {et~al.} 2009, \apj, 692,
  187

\bibitem[{{Wolfire} {et~al.}(1995){Wolfire}, {Hollenbach}, {McKee}, {Tielens},
  \& {Bakes}}]{wolfire95}
{Wolfire}, M.~G., {Hollenbach}, D., {McKee}, C.~F., {Tielens}, A.~G.~G.~M., \&
  {Bakes}, E.~L.~O. 1995, \apj, 443, 152

\bibitem[{{Wolfire} {et~al.}(2003){Wolfire}, {McKee}, {Hollenbach}, \&
  {Tielens}}]{wolfire03}
{Wolfire}, M.~G., {McKee}, C.~F., {Hollenbach}, D., \& {Tielens}, A.~G.~G.~M.
  2003, \apj, 587, 278

\bibitem[{{Zhang} {et~al.}(2014){Zhang}, {Thompson}, {Murray}, \&
  {Quataert}}]{zhang14}
{Zhang}, D., {Thompson}, T.~A., {Murray}, N., \& {Quataert}, E. 2014, \apj,
  784, 93

\end{thebibliography}

\end{document}